\documentclass[traditabstract]{aa} 

\usepackage{natbib,graphicx}
\usepackage{txfonts}
\begin{document}
\title{Stellar populations in the Milky Way bulge region: Towards solving the Galactic bulge and bar shapes using 2MASS data}

   \author{Annie C. Robin
          \inst{1}
          \and
          Douglas J. Marshall\inst{2,3}
          \and
           Mathias Schultheis\inst{1}
        \and
        C\'eline Reyl\'e \inst{1}
          }

   \institute{Institut Utinam, CNRS UMR6213, OSU THETA, Universit\'e de Franche-Comt\'e, 41bis avenue de l'Observatoire, 25000 Besan\c{c}on, France   \\
              \email{annie.robin@obs-besancon.fr}       
\and Universit\'e de Toulouse, UPS-OMP, IRAP, F-31028 Toulouse cedex 4, France
\and CNRS, Institut de Recherche en Astrophysique et Plan\'etologie, 9 Av. colonel Roche, BP 44346, F-31028 Toulouse cedex 4, France}

   \date{}


  \abstract
   {Exploring the bulge region of our Galaxy is an interesting but challenging quest because of its complex structure and the highly variable extinction. We re-analyse photometric near-infrared data in order to investigate why it is so hard to reach a consensus on the shape and density law of the bulge, as witnessed in the literature. The apparent orientation of the bulge seems to vary with the range of longitude, latitude, and the population considered. To solve the problem we have used the Besancon galaxy model to provide a scheme for parameter fitting of the structural characteristics of the bulge region. The fitting process allows the shape of the bulge's main structure to be determined . 
   
We explore various parameters and shapes for the bulge population, based on Ferrer's ellipsoids, and fit the shape of the inner disc in the same process. The results show that the main structure has a standard triaxial boxy shape with an orientation of about 13$^{\circ}$ with respect to the Sun-Galactic centre direction. But the fit is greatly improved when we add a second structure, { which is a longer and thicker ellipsoid. We emphasize that our first ellipsoid represents the main boxy bar of the Galaxy and that the thick bulge population could be either (i) a classical bulge slightly flattened by the effect of the bar's potential, or (ii)  an inner thick disc counterpart}.

 { With Ferrer's ellipsoid, the model shows a general agreement with 2MASS data at the level of 10\% in the whole bulge region but does not produce the ``double clump'' feature. However, we show that the double clump seen at intermediate latitudes can be reproduced by adding a slight flare to the bar. To characterize the populations better, we further simulate several fields that have been surveyed in spectroscopy and for which a metallicity distribution function (MDF) are available. The model agrees well with these MDF measured along the minor axis if we assume that the main bar has a mean solar metallicity and the second thicker population has a lower metallicity. It then naturally creates  a vertical metallicity gradient by mixing the two populations.}

{ In the process of model fitting, we also determine the thin disc parameters. }The thin disc is found to have a scale length of 2.2 kpc, in good agreement with previous estimates towards the anticentre, but with a large hole of scale length 1.3 kpc, giving a maximum density in the plane for this population at about 2.3 kpc from the Galactic centre.
In the very central part of the bulge on top of our two populations, and by subtracting the fitted ellipsoids, we find evidence of an extra population in the nuclear region, at 2$^{\circ}$ in longitude and 1$^{\circ}$ of latitude from the Galactic centre. Its location corresponds well to the central molecular zone, and to the Alard nuclear bar.


   \keywords{     Galaxy: stellar content, Galaxy: evolution, Galaxy: bulge, Galaxy: formation, Galaxy: dynamics          }
}
\titlerunning{The Galactic bulge/bar shape}
   \maketitle


\newcommand{\Hbeta}{H$\beta$ }
\newcommand{\cpara}{$c_{\parallel}~$}
\newcommand{\cperp}{$c_{\perp}~$}
\newcommand{\Msun}{$M_\odot~$}
\renewcommand{\deg}{$^{\circ}$}
\newcommand{\Ro}{$R_{\odot}~$}

\section{Introduction }

Since the discovery of a triaxial structure in the central regions of the Galaxy {from 24 micron observations \citep{1991ApJ...379..631B}, COBE NIR data \citep{Weiland94,Dwek}, or visible photometry \citep{1994ApJ...429L..73S}}, numerous attempts have been made to characterize this structure and to investigate its origin. It is still unclear whether this structure, often called the outer bulge but sometimes the bar, had its origin in the early formation of the spheroid (like a typical bulge, similar to ellipsoidal galaxies) or was formed by a bar instability later in the disc. It is crucial to investigate this formation history as a benchmark in understanding formation of disc galaxies. Kinematical studies are very important for tracing the stellar motions and deducing a dynamical history of the populations in the inner regions. New surveys in the near future will help solve this question, particularly the Gaia mission of the European Space Agency. 

In the meantime photometric surveys, especially in the infrared, can help to determine the shape and density laws of the stellar populations in place. The question of the shape, orientation and extent is still open, because one can find   for the angle of the major axis direction with regard to the Sun-Galactic centre direction values between 10\deg\  and 45\deg, depending on the fields considered, the wavelength and the method used. Among all the attempts, the significant ones include \cite{1994ApJ...429L..73S}, \cite{Dwek}, \cite{Fux97}, \cite{Lopez2000}, \cite{Freudenreich}, \cite{Picaud04}, \cite{Babusiaux2005}, \cite{Benjamin}, and\cite{Rattenbury}. The difficulty in accurately measuring the orientation of the bulge might be aggravated if, actually, there is more than one population present in the region and if these populations have different origins.  { \cite{Martinez2011} also offer a clever explanation for the lack of consensus concerning the bar angle, which they think is partlyfrom the cone effect and partly from the presence of leading or trailing overdensities at the end of the bar, as also shown by \cite{Romero2011}. 

In this context, metallicity distribution studies can be useful constraints for distinguishing populations. \cite{Zoccali2008} and \cite{Hill2011} highlight a bimodal metallicity distribution in Baade's window and several fields on the bulge minor axis, thus identifying a mixture of populations}. Concerning the kinematics, proper motions are difficult to obtain due to the large distance of the bulge, while radial velocities are expensive to process for large numbers of stars. { \cite{Sumi} studied proper motions from the OGLEIII survey and compared them with a dynamical model. However, the OGLE fields avoid the Galactic plane and the northern part of the bulge, they are limited by extinction(because the observations were performed in the visible), and the mean proper motion error is too large to accurately determine velocities at the bulge distance}. 

\cite{Rich2007} have undertaken a wide survey (BRAVA, for bulge radial velocity assay) of radial velocities at longitudes between -10\deg\  and 10\deg\  and at latitudes -4\deg\  and -8\deg. Their first results from the latitude -4\deg\  data set show that the slope of the rotation curve flattens considerably at longitude $ |l|>5$\deg\  indicating a probable change in the dominant population at this point. {Furthermore, the analysis of metallicity versus kinematics by \cite{Babusiaux2010} for RGB stars in three bulge fields, at different latitudes along the minor axis, show that two populations with distinct chemo-dynamical signatures seem to be present in these fields. 
Since 2010, there have been several studies showing double clumps in a number of fields in the bulge \citep{Nataf10,Mcwilliam10,Saito11}. This is also a feature that should be taken into account in an overall view of the bulge region and that the models should explain.

Complete understanding of the bulge's structure and formation would require studying the abundances and kinematics in various fields at different longitudes and latitudes,such as the APOGEE project \citep{Majewski}. In the mean time we propose to explore the bulge structure based on an analysis of  }
 existing photometric data. We intend to check whether a detailed modelling of the stellar populations and interstellar extinction present in the central region might lead to an answer. We use the Two Micron All Sky Survey, hereafter 2MASS \citep{Skrutskie} photometry and a model-fitting procedure based on the Besan\c{c}on Galaxy model  (BGM) \citep{Robin03} and explore a wide region -20\deg$<l<20$\deg\  and -10\deg$<b<10$\deg\  to fit a multi-dimensional space of parameters defining the density laws of the bulge and disc populations. 
{ The analysis proceeds in several steps. First we attempt to fit the whole area with a Monte Carlo scheme to explore the parameter space using the differential star counts in bins of K and J-K in a set of selected windows. We compare the resulting model with integrated star counts in order to visually check the realism of the resulting model. In a second step we compare simulated CMDs in a sample of directions to check their agreement in more detail. At this stage we introduce a modulation of the density distribution of the bar structure to reproduce the double clump structure seen in medium latitude directions by \cite{Nataf10}, \cite{Mcwilliam10}, and \cite{Saito11}. In a final step we compare model simulations of the metallicity distribution functions with data over the minor axis of the bar, placing constraints on the mean metallicity of the composite populations in the bulge region. }
 
In section~2 we describe the basis of the BGM and the free parameters which we adjust to the existing data. The data are described in section~3 along with the method that corrects for the 3D extinction distribution towards the bulge. The fitting method and results are given in section~4. Validation of our extinction method is provided in section~5. In section~6, we compare model predictions with CMDs in several directions and explain how the model can reproduce the double clump feature in the CMDs. In sect. 7, we zoom in the internal bulge.  In sect. 8 we discuss the results, and the metallicity distribution function, and present the perspectives of this work. 

\section{Population synthesis model for the bulge}

The stellar population model is based on a population synthesis scheme. Four distinct
populations are assumed (a thin disc, a thick disc, a bulge, and a spheroid),
each deserving specific treatment. In the central regions, only
the bulge and thin disc are important. The thick disc is a minor component, and the halo completely
negligible at the magnitudes and latitudes considered here. The thick disc contributes slightly to the star density but only when the bulge and the thin disc become less important, that is, at latitudes above about 8-10\deg.

\subsection{The thin disc population}

The thin disc contributes significantly to star counts in the Galactic
central region, so its characteristics have to be fitted at the same time
as the bulge parameters. The important parameters are the scale length and the size of the central hole. Other parameters and structures, such as outer radius, warp, and flare, are
taken into account in the BGM and described in \cite{Reyle09}, but are not relevant here. 

A standard evolution model is used to produce the disc population,
based on a set of evolutionary tracks,
a constant star formation rate (hereafter SFR) over 10 Gyr,
and a two-slope initial mass function 
$\phi(M)=A\times \cdot M^{-\alpha}$ with $\alpha$=1.6 for M$< 1 M_\odot$ and
$\alpha$=3.0 for M$>$1$M_\odot$.
The preliminary tuning of the disc evolution parameters against relevant 
observational data was described in \cite{1997A&A...320..428H,1997A&A...320..440H}
and later changes are explained in \cite{Robin03}.

       With the evolution model we populate the thin disc dividing it into seven age components. 
For each subcomponent, the distribution in absolute magnitude and effective temperature is obtained, assuming a star formation history constant between 0 and 10 Gyr. The star counts are computed using the standard equation of stellar statistics from the distribution in $M_{V}$ and a spatial density distribution. It is therefor equivalent to assuming that the SFR is roughly the same over all the thin disc.

        The thin disc density distribution model follows the
\cite{Einasto79} law: the distribution of each disc component (except for the very young one of age less than 150 million years, which is not important here) is
described by an axisymmetric ellipsoid with an axis ratio depending on
the age. The density law of the ellipsoid is
described by the subtraction of two functions:
$$\rho_d = \rho_{d_0} \times [\exp (-\sqrt{0.25+
(\frac{a}{R_d})^2})
-\exp (-\sqrt{0.25+(\frac{a}{R_h})^2})]$$
\begin{center}
with $a^2=R^2+\left(\frac{Z}{\epsilon}\right)^2$, where:
\end{center}
\begin{itemize}
\item $R$ and $Z$ are the cylindrical Galactocentric coordinates;
\item   $\epsilon$ is the axis ratio of the ellipsoid. Values of $\epsilon$ as a function of age and local normalization are given in \cite{Robin03};
     \item   $R_d$ is the scale length of the disc and is around 2.2-2.5 kpc
(Ruphy et al. 1996). This parameter will be fitted in the procedure described in this paper;
\item   $R_h$ is the scale length of the hole. It is also fitted here.
 The maximum density of the disc population is approximately at 2 kpc from the centre;
\item The normalization $\rho_{d_0}$ is deduced from the local luminosity
function (Jahrei{\ss} et al, private communication), assuming that the Sun is
located at R$_\odot$=8 kpc and Z$_\odot$=15 pc. Alternative values for the sun-Galactocentre distance R$_\odot$ are also considered.
\end{itemize}

\subsection{The outer bulge}

\cite{Picaud04} undertook a detailed analysis of the outer bulge stellar
density and luminosity function by fitting model parameters to a set
of 94 windows in the outer bulge situated at -8$^\circ<l<10^\circ$ and 
-4$^\circ<b<4^\circ$. The data were obtained by the
DENIS survey team \citep{Epchtein} using K$_{\rm s}$ magnitude
and J-K$_{\rm s}$ colour distributions. Using a
Monte Carlo method to explore an 11-dimensional space of bulge and disc 
density model parameters
and a maximum likelihood test, they showed that the
bulge follows a boxy exponential or 
a boxy sech$^2$ profile like the one fitted to the {\it DIRBE} integrated flux in the near infrared \citep{Freudenreich}. These are Ferrers ellipsoids \citep{Ferrers}. 
Their resulting triaxial bulge has a major axis
pointing towards the first quadrant with an angle of about 10$^\circ$ 
with respect to the Sun-Galactic centre direction. A full description of the  parameter 
values of the bulge density law can be found in \cite{Picaud04}. 

Here we reconsider the density law fitted by \cite{Picaud04}. 
The Freudenreich S function is:
\[
 \rho_S = \rho_0 ~ \mbox{sech}^2 (-R_s) 
\]

with 
\[
R_s^{C_\parallel}=[|\frac{X}{x_0}|^{C_\perp}+
|\frac{Y}{y_0}|^{C_\perp}]^{C_\parallel/C_\perp}+
|\frac{Z}{z_0}|^{C_\parallel}\\
\]
(X,Y, Z) are the cartesian coordinates in the referential of the triaxial structure (X being the major axis, Y the second axis and the Z the third).
The parameters \cpara and \cperp are important for exploring a wide range of shapes, from ``disky'' to ``boxy''. This allows great flexibility: one can even have a ``disky'' shape in the plane, together with a boxy projection vertically. 

In the following we also consider alternative functions for the overall shape, exponential (E), and Gaussian (G):

\[
 \rho_E = \rho_0 ~\mbox{exp} (-R_s)  \]

and 

\[
 \rho_G = \rho_0  ~\mbox{exp} (-R_s)^{2}. 
\]

The density function is then multiplied by the cut-off function
$f_c$ (distances given in kpc, and $R_c$ is called the cut-off radius):

\begin{center}
\begin{tabular}{rl}
$\rho=\rho ~ f_c(R_{XY})$, & \mbox{with} $R_{XY}=\sqrt{X^2+Y^2}$\\
$R_{XY} \leq R_c$ $\Longrightarrow$ & $f_c(R_{XY})=1$\\
$R_{XY} \geq R_c$ $\Longrightarrow$ &
$f_c(R_{XY})=\exp \left(-(\frac{R_{XY}-R_c}{0.5})^2\right)$\\
\end{tabular}
\end{center}

Three angles define the orientation:
\begin{itemize}
\item   $\phi$: orientation angle
from the sun--centre direction of the projection
on the Galactic plane of the bulge major axis,
\item   $\beta$: pitch angle of the bulge major axis from the Galactic
plane,
\item   $\gamma$: roll angle around the bulge major axis.
\end{itemize}

 In the  \cite{Picaud04} study, the angle $\beta$ was found 
very close to 0$^\circ$, and the third angle $\gamma$ was found to be ill defined because
the minor axes Y and Z had similar scale lengths.
In this study, we therefore adopt $\beta$=$\gamma$=0$^\circ$. Finally, the angle $\phi$, the three scale lengths
$x_0$, $y_0$, $z_0$,
the density at the centre $\rho_0$, the cut-off radius $R_c$, and
the two coefficients $C_\parallel$ and $C_\perp$ are considered
as free density parameters in the fitting process.

The bulge stellar population is taken from a single burst population of ages varying between 6 and 10 Gyr as tested by  \cite{Picaud04}. The favoured combinations
between the age and evolution models  are
from \cite{Bruzual1997} with an age of 10 Gyr, or from
\cite{Girardi} with an age of 7.94 Gyr (log(age)=9.9). It appears that the K$_{\rm s}$ band counts are not very sensitive to the age, so a proper estimation of the age and age range of the populations in the bulge would require complementary data. This analysis thus takes the luminosity function from \cite{Girardi} with a log(age) of 9.9 as a reference.

\subsection{One or two populations in the bulge?}

In the bulge region there may be several populations, with several star formation histories, cohabiting.
The coexistence of different populations in the bulge region can be tested using the distribution in colour-magnitude diagrams. In our fitting process we consider alternatively one or two bulge populations, and for each of them the S, E, or G functions.

\section{Data description and extinction corrections}

\subsection{2MASS data}

The \cite{Picaud04} analysis was limited in longitude and latitude (-12\deg$<l<8$\deg, -4\deg\ $ <b<$4\deg), used DENIS data, and did not contain many fields near the plane (less than 10\% were situated at $|l|<1$\deg). Here,
we use the 2MASS point source catalogue data. We concentrate on the region -20\deg\ $<l<$20\deg, -10\deg\ $ <b<$10\deg. Unlike the \cite{Picaud04} study, the selection goes down to latitude b=0\deg. The data are used in three ways: (i) for determining the extinction using the method from \cite{Marshall06}, (ii) for fitting bulge model parameters on a selection of windows, (iii) for comparing the whole bulge region with the fitted model and checking the extinction distribution again.

\subsection{Extinction}
Since the extinction has a major effect on star counts close to the plane, a detailed analysis of the extinction field by field is needed to provide realistic model simulations. For this purpose, we used
the method described in \cite{Marshall06}. This method adopts the standard BGM and determines the extinction by comparing the simulated colour-magnitude diagrams from this model and the observational data. 
Adjusting both the extinction and the bulge parameters in a single run is a difficult data analysis problem. Therefore we proceeded with iterations, first fitting the extinction with the preliminary bulge model \citep[from][]{Picaud04}, then adjusting the bulge parameters assuming this extinction and finally verifying that the extinction derived from the new fitted bulge model is not significantly different from the original extinction distribution. 

The goodness of fit used in the extinction determination is based on a chi-squared statistic for two binned distributions having different total 
number of elements \citep{Press92}. As such, the extinction determination is sensitive to differences in colour and not to discrepancies between the 
number of observed and modelled stars. 

In the present study, the K$_{\rm s}$ magnitude and the J-K$_{\rm s}$ colour were used to
compare with simulations. Cuts were made in K$_{\rm s}$ and J at the completeness limit, defined from the histogram in magnitude at the bin before the maximum for limiting the risk of incompleteness that can occur even at the peak in magnitude. In the extinction determination and in the maps for visualizing the comparison between model and data (see section 4.1 and figures 2, 3, and 4) we further restrict the comparison at K$_{\rm s}<$12 because the extinction determination method is only based on the giants' colour and that they can be separated from the dwarfs only up to this magnitude.

\section{Fitting procedure}

\begin{figure}
   \centering
\includegraphics[width=6cm,angle=-90]{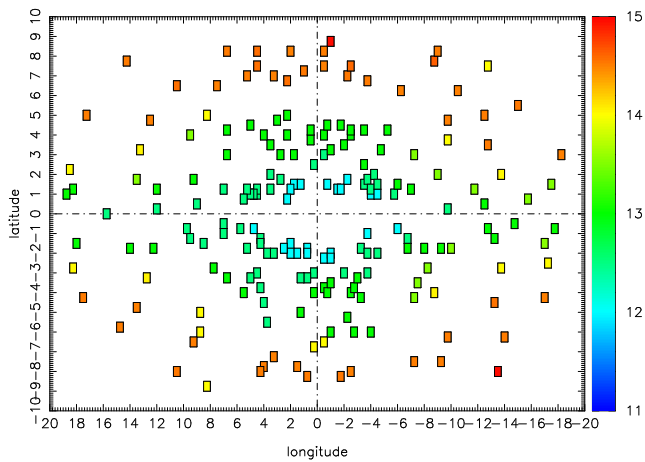}

      \caption{Distribution in longitude and latitude of the 200 windows used in the fitting process. The colour coding corresponds to the limiting magnitude of the counts in K$_{\rm s}$.}
       \label{windows}
   \end{figure}

The first set of data we use is a selection of fields from the 2MASS point source catalogue. Two hundred  windows have been chosen in order to cover the region defined by -20\deg$<l<$20\deg and -10\deg$<b<$10\deg. For each window, the extinction is determined using the method described in section 3.2, as well as the completeness limit  independently in each band. The BGM in its standard version is used to simulate the NIR data, taking the completeness into account. To illustrate, we show in fig.~\ref{windows} the 200 windows and the completeness limit in the K$_{\rm s}$ band. Then the following procedure  is used to adjust the bulge model parameters. 

For each window the star counts in the CMD diagram are binned { in K$_{\rm s}$ and J-$K_{\rm s}$ in bins with equal numbers of stars, two bins in colour, and seven  in magnitude, giving a total of 14 bins. In the model the counts of each population are counted separately to be able to change them as a function of the variable parameters}. 
The parameter space is explored randomly from successive Monte Carlo drawings and a selection of the best models in the sense for the likelihood. The likelihood function is the same as in \cite{Picaud04}. The process is stopped when the variation in the parameters from one iteration to the next is less than the required accuracy for each parameter. The process is done 20 to 50 times to ensure a robust solution.

In a second step, we compare the fitted model with 2MASS data in the whole region $-20$\deg$ < l <20$\deg and $-10$\deg$< b < 10$\deg\  and compute the relative difference between data and model on a grid of 15$\times$15 arcmin fields. The aim is twofold. Firstly, it ensures that the 2MASS data are correctly fitted by the model all over the bulge area (not only in the 200 selected windows), without systematics in any region. Secondly, it verifies that the new extinction, recomputed from the new density model, has not varied significantly since the density fitting process, so we ensure that proceeding in two separate steps (fitting of the extinction, then fitting of the density model) is efficient and will converge.

We first attempted to fit a unique bulge population all over the considered region. The parameters that are fitted are the following: the bulge angle $\phi$ (angle between the major axis of the bulge and the Sun-Galactic centre direction), the bulge scale lengths (considering 3 axes, x$_{0}$, y$_{0}$, z$_{0}$), the bulge mass, the extent in radial distance (cutoff R$_{c}$), two parameters describing the boxiness of the bulge \cpara and \cperp (see above), the disc scale length, and the scale length describing the size of the hole inside the disc.

As we show below, the agreement was not good enough, and we had to consider alternative models. 
We attempted to fit a second structure, with the same kinds of parameters as for the first bulge structure. We also considered different formulae for the bulge density and for the second structure: sech$^2$, Gaussian, or exponential. 

We assume here a Sun to Galactocentric distance of 8 kpc (following recent results from \cite{Ghez}),  unlike  \cite{Picaud04} who assume a value of 8.5 kpc. We investigate the effect of this value on the result in the discussion.

In the next subsections we describe the results for each kind of model fitted.

\subsection{Fitting a single ellipsoid}

The fit of a single ellipsoid on the large area considered here leads to significantly different results compared with the analysis done by \cite{Picaud04} where the data set considered was restricted to the region  $-12$\deg$<l<8$\deg\  and $-4$\deg$<b<4$\deg\  with about 100 windows. The important thing here is probably not the number of windows, which is doubled, but rather the extent of the area. { Moreover, unlike the \cite{Picaud04} study, we drop the selection of giants and consider a fit covering the whole colour range. Finally, we account for interstellar extinction in a more realistic manner than they did.}

\begin{table*}
\label{param_1pop}
\caption{Best fit models with 1 single bulge ellipsoid. The columns give the fitted parameters and reduced likelihood Lr. $\phi$: bulge angle of the major axis with the Sun-Centre distance in degrees; x0, y0, z0: scale lengths along the 3 axes in kpc; normalization factor; Rc: cutoff radius in kpc, \cpara and \cperp: parameters of boxyness, Rd: disc scale length in kpc, Rh: disc hole scale length in kpc. }

\begin{center}
\begin{tabular}{llllllllllllll}
\hline
Model & $\phi$ & x0 & y0 & z0 & Normalization & Rc &  \cpara & \cperp & Rd & Rh & Lr \\
  & \deg & kpc & kpc & kpc & 1.e9 \Msun pc$^{-3}$ & kpc &  & & kpc &kpc & &  \\

\hline
A & 11.10 & 1.59 & 0.424 & 0.424 & 9.63 & 2.54 &  2. & 2. & 2.53 & 1.32 & - \\

\hline
\\
B &    7.1 & 4.07 & 0.76 & 0.41&  23.83 &  5.99  &  1.434  &3.797 & 2.26 & 0.18 & -22842.\\ 
stdev  &   1.01  &  0.263  & 0.043 & 0.018  & 1.31 & 0.025 & 0.162 & 0.858 &0.165 & 0.190 & 892.\\
\\
C & {\it 20.0} & 4.48 & 1.46 & 0.47 &26.47 &  5.98 &  1.015 & 4.579 & 2.64 & 0.99 &  -32895.\\ 
stdev & 0 & 0.403 & 0.122 & 0.023 & 0.928 &0.0188 & 0.135 & 0.491 & 0.302 & 0.183 & 1707. \\
 \\
D &  \it{25.0} & 3.77 & 1.48 & 0.46 &29.09&  6.00 &   1.019   &4.379 & 2.47 & 1.03 & -40027.\\ 
 stdev& 0 & 0.325 & 0.084 & 0.024 & 1.09 & 0.022 & 0.123 & 0.542 & 0.321 & 0.283 & 1578. \\
 \\
G &   5.8 & 4.12 & 0.84 & 0.41 & 16.43& 5.99 &  1.271&  2.223 &  2.30 & 0.61 & -22943. \\
stdev & 1.12 & 0.260 & 0.121 & 0.016 & 0.273 & 0.039 & 0.197 & 0.870 & 0.208 & 0.250 & 609. \\
 \\
E &    6.9 & 2.54 & 0.45 & 0.23 & 49.75 &  5.98&  1.476 & 4.656 &  2.25 & 0.60 &  -23661. \\
stdev & 1.21 & 0.310 & 0.050 & 0.013 & 2.85 	& 0.083 & 0.381 & 0.977 & 0.167 & 0.247 & 1514. \\

\hline

\end{tabular}
\end{center}
\end{table*}

Results are given in Table~1. Model A indicates the parameters fitted on DENIS data by \cite{Picaud04} for comparison only. In the model A fitting process, the parameters \cpara and \cperp were fixed to 2 and the Sun-galactocentric radius to 8.5 kpc. For each model and each parameter, the dispersion around the fitted values is indicated in the second line. When a model has a parameter fixed rather than fitted, it is in italics. The B model is the model with a free angle with the (default) sech$^2$ shape. C is similar but with a fixed angle of 20\deg, D is similar but with a fixed angle of 25\deg. The G model is similar to the B model but with a Gaussian shape, the E model has an exponential shape (see below). 

 The most significant difference between the new result (model B) and the old one (model A) is the new scale length of the bulge (major axis scale length x$_{0}$), which was found to be 
1.59 kpc in \cite{Picaud04} and is here about 4 kpc. The cutoff radius has also grown from 2.54 to about 6 kpc. It means that the limited region used in our first study has led to underestimating of the bulge's extent.
The second axis scale length is also greater than in \cite{Picaud04} (corresponding to the depth
of the bulge along the line of sight if the angle is small), while the scale height (minor axis z$_{0}$) has not changed  (the latitude range covered by the previous paper was sufficient to cover the bulge). 
We also found that the scale length of the disc has slightly diminished and that the hole has
diminished. 

\begin{figure*}
   \centering

   \includegraphics[width=7cm]{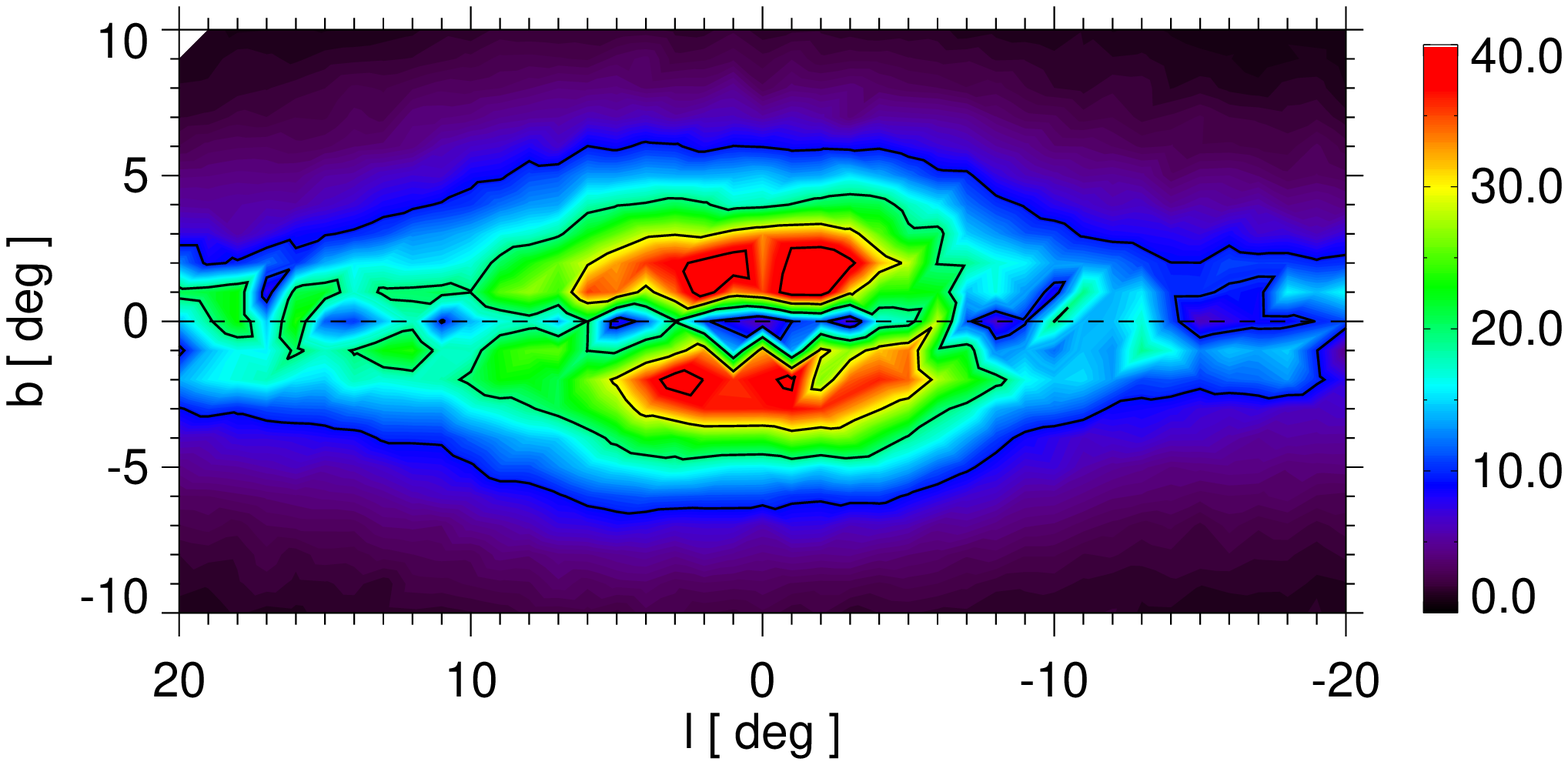}
    \includegraphics[width=7cm]{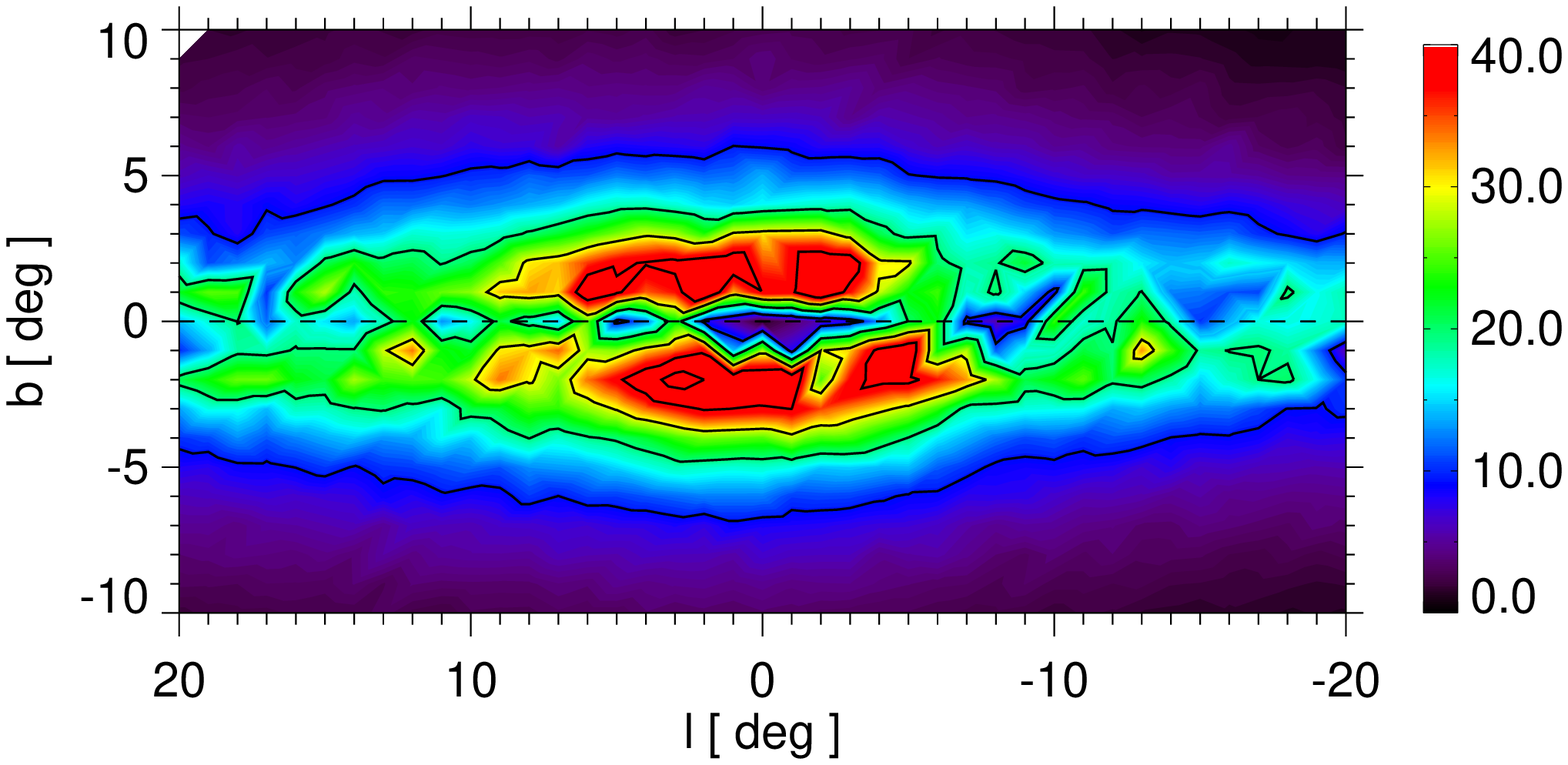}
     \includegraphics[width=7cm]{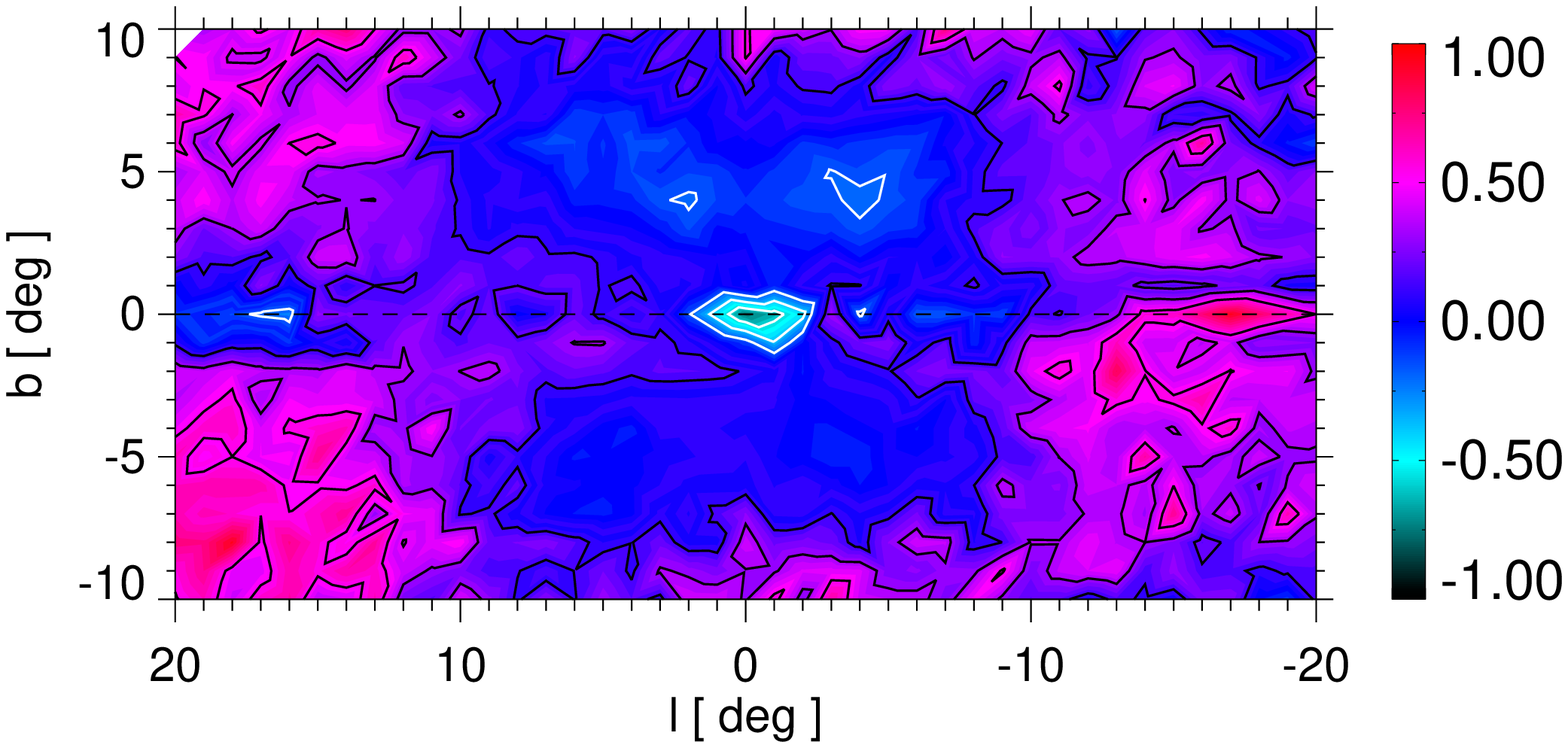}
   \includegraphics[width=7cm]{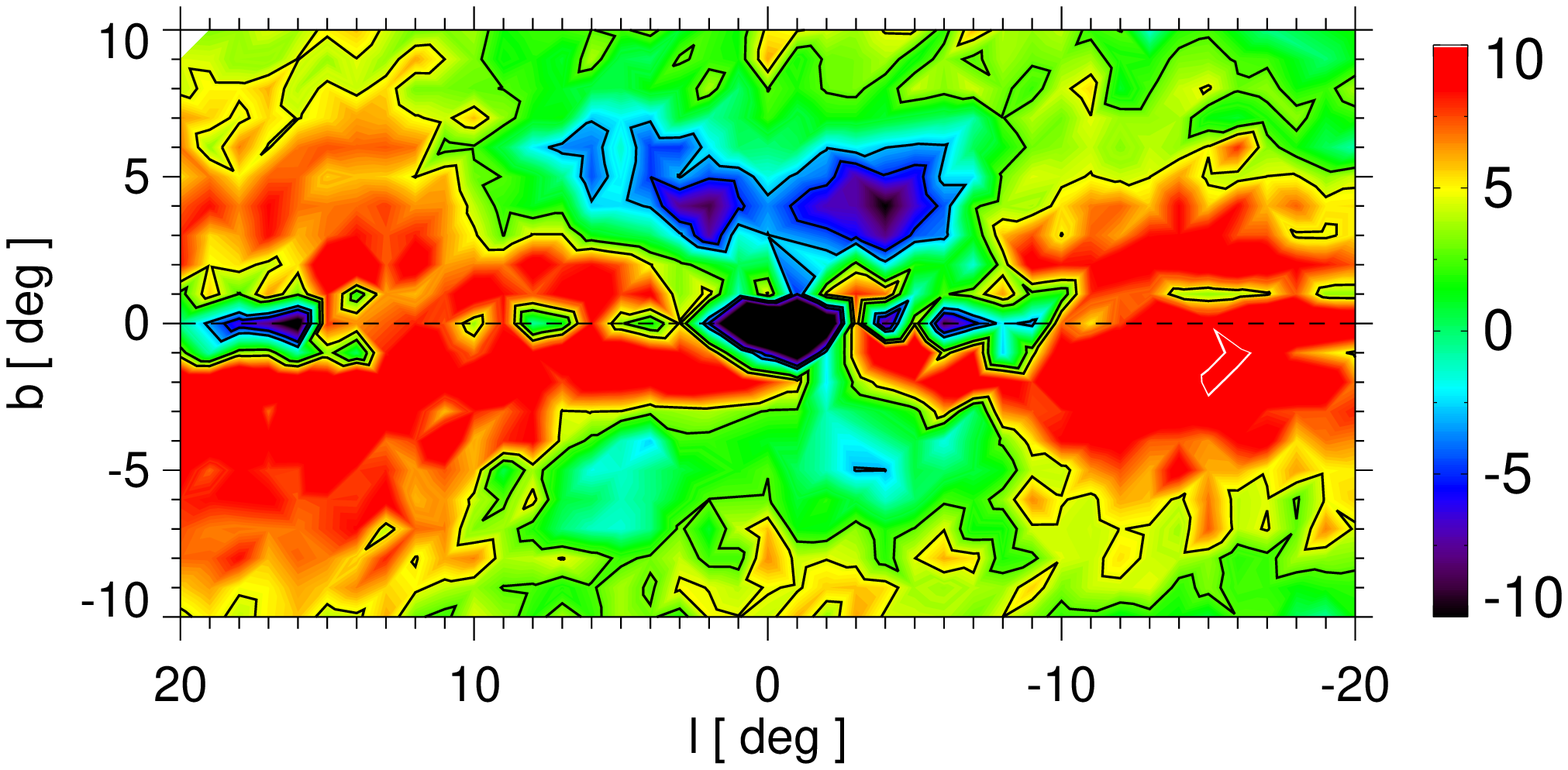}
      \caption{Star counts up to limiting magnitude K$_{\rm s}$ from 2MASS data (top left) compared with fitted 1 ellipsoid model B (top right), residuals (N$_{\rm mod}$-N$_{\rm obs}$)/ N$_{\rm obs}$ (bottom left), and difference divided by the Poissonian counting error of the model (i.e. (N$_{\rm obs}$-N$_{\rm mod}$)/ $\sqrt(N_{\rm mod})$ (bottom right). In the residual map, contours are drawn at intervals of 20\% model overestimate (black) and 20\% model underestimate (white).
          The 1-ellipsoid model leaves significant X-shaped residuals. Near the Galactic centre the nuclear bar population is missing in the model. The residuals in the outer regions are not very significant due to the small number of stars in each bin (seen in the top left panel in dark violet and black).}
       \label{1pop}
   \end{figure*}

We plot in figure~\ref{1pop} star counts in K$_{\rm s}$ from 2MASS data (top left) compared with the fitted model with single population (top right), residuals as defined by $(N_{\rm mod}-N_{\rm obs})/N_{\rm obs}$ (bottom left), and  difference divided by the Poissonian counting error of the model (i.e. (N$_{\rm obs}$-N$_{\rm mod}$)/ $\sqrt(N_{\rm mod})$ (bottom right). 
When computing the extinction, the K$_{\rm s}$ band was limited to 12 or brighter. The other bands have no such limitation. This limit ensures lower uncertainty on the 2MASS PSC fluxes.
This figure shows the B model described in Table~1 after the 3D extinction distribution has been refitted. 
 The structure of the residuals shows that the shape of the fitted bulge model is not appropriate. In the in-plane region the model density agrees well with the data. But the fields at high latitudes and intermediate longitudes have X-shape residuals, where the model density is too low. It is probable that the resulting fit is a bad compromise between adjusting the densities in plane and at different latitudes. It is worth noting that in the top left-hand  panel the 2MASS star count map indeed shows boxy iso-densities that are not reproduced by the model, even if the density laws were chosen explicitly to allow for a boxy shape. 

Interestingly, the residual map shows a structure in the inner bulge at longitudes less than 2\deg, which we believe is the nuclear bar seen by \cite{Alard}. This population has not yet been included in the model, and is shown here by subtraction. We discuss the point in section 7.

\subsection{Influence of other parameters}

The Sun-Galactocentric distance \Ro is not adjusted here during the fitting process. It is fixed at either 7., 7.5, 8, or 8.5 kpc. Models with \Ro of 8 kpc give the larger likelihood, but the data are not sensitive enough to distances along the line of sight to strongly constrain this parameter. Moreover, changing  \Ro has little impact on the other parameters, such as \cpara
and \cperp, which stay of the order of 1 and 4 (resp). 

We assume an age of about 8 Gyr for the bulge population (log(age)=9.9), following the best result in the \cite{Picaud04} study. 
{Our analysis is not very sensitive to the assumed age for an old population (age over 5 Gyr) in the red clump. The difference in the luminosity function would be noticeable if we had covered the red clump and the giant branch down to the turnoff of the population, which is not possible with 2MASS data.}
The turnoff position and the giant sequence are also influenced by the age. The turnoff would be slightly brighter in a younger population (but this is not visible in most 2MASS fields), and the giant sequence would be slightly bluer. But this effect is of second order and difficult to see in broad band photometry and even more in extincted fields. Complementary data would be necessary to constrain the age of the population at the level of a Gyr. Consequently, we consider an isochrone for the bulge with an age of about 8 Gyr from \cite{Girardi}, which should be valid if the true age of the population is in the range 6-10 Gyr. Alternative isochrones within a reasonable range of age would not significantly change the conclusion on the bulge shape and the extinction. 

\subsection{About the bulge/bar angle}

Many bulge studies find a significantly larger bulge angle than our result (from one population model B : 7.1\deg). Most of them are in the range 10\deg\  to 45\deg. We found 11.1\deg\ in model A. Since some of our parameters are slightly correlated, we attempted to fix the bulge angle at different values to see whether it is possible to find a good  enough fit with a larger bulge angle and to see in which field this value is favoured. Models C and D  in Table~1  have  fixed bulges angle of 20\deg\  and 25\deg\,  respectively. These models have very low likelihood with regards to smaller angles.

We suspect that the bulge angle found in different studies might depend on the field selection, extent in longitude, distance to the Galactic plane, and depth of the survey. \cite{Martinez2011} have shown how the bar angle determination can be strongly biased by the cone volume effect, in the case of the long bar, but this effect is also present when determining  the main bulge angle using the mean distance of the red clump.

To test these effects, we selected subsets of fields in longitude and latitude and reproduced the fitting procedure. It appears that for fields at $|l|>12 $\deg\   
and close to the plane ($|b|<3$\deg), the likelihood is higher with a large angle, while this large angle
gives a bad fit at low longitude and high latitude ($|b|>3 $\deg\  and $|l|<5 $\deg). 
Fields at high longitudes and low latitudes are well fitted by large bulge angles, while fields at low longitudes and high latitudes favour small bulge angles. When fixing the angle of the bar at higher values (models C and D, 20 or 25\deg), the fields situated at large longitudes show similar likelihood, but fields at higher latitudes ($|b|>3 $\deg) get significantly worse. 
This is in good agreement with the analysis from \cite{Cabrera07,Cabrera08}, from 2MASS data as well, but limited to fewer fields. These findings might indicate that a single ellipsoid fit is not adequate. This is why in the following section we attempt to fit the sum of two ellipsoids on our data set.

\subsection{Fitting two ellipsoids}

We attempt to apply the procedure to fit two ellipsoids (in addition to the thin disc) having the same luminosity function, age, and metallicity, as these parameters do not influence the overall near infrared photometry very much (see the discussion below). The same parameter set describing each ellipsoid is fitted on both populations. We consider several combinations of shapes, S shape as before (S), exponential (E), Gaussian (G) as in \cite{Dwek}. 

Table~2 shows the parameters of the best fit with different shapes S, E, or G. Because it is unlikely that two populations have different angles from a dynamical point of view, and because the fit with different angles did not give significantly different angles, we  have imposed the two ellipsoids to have the same orientation in the plane. 

\begin{table*}
\label{2aligned}
\caption{Parameters and reduced likelihood Lr when two ellipsoids are fitted, including main sequence and giants, and forcing the two ellipsoids at the same orientation. Parameters are the same as in previous table. The last two lines of the table give the standard deviation about the mean for model (S+E). Standard deviation for other models are similar.}
\begin{center}
\begin{tabular}{lllllllllllllll}
\hline
Shape & N & $\phi$ & x0 & y0 & z0 & Normalization & Rc & \cpara & \cperp & Rd & Rh &Lr \\
 & & \deg & kpc & kpc & kpc &1.e9 \Msun pc$^{-3}$ & kpc & & & kpc & kpc &  \\

G+G & (1)  &  12.5 & 1.63 & 0.51 & 0.39 & 20.71 & 2.67 &  3.040 & 2.224& 2.36  &1.31 &  -14059. \\
 & (2) &  - &4.27 & 1.32 & 1.18 &1.58 &  6.93  & 3.509 & 3.898 &   & &\\
 \\
 G+S & (1) &  10.6 & 1.49  &0.52 & 0.40 & 18.20 & 3.28 &   3.102 & 3.178  &2.26 &  1.53  & -14770. \\
  & (2) & - &  4.57 & 1.45 & 0.92& 2.17 & 6.80  & 3.699&  3.300&   & & \\    
 \\
 G+E &(1) &  10.0 & 1.36 & 0.52 & 0.38&  22.53 & 3.28  &  4.318 &  1.962  &2.45 & 1.35 & -13417. \\
 & (2) & - &  3.98  &1.33 & 1.00 & 2.59 &  6.91 &  4.201 & 2.447 &  &&\\   
\\
S+S & (1) &  11.7  &1.47 & 0.48 & 0.39 &32.79 &  2.91&  3.727  &3.326  &  2.29 & 1.56 & -14134. \\
& (2) & - &  4.69 & 1.44 & 1.38& 2.10 & 6.75 &  2.437 & 4.524 & & &\\  
\\
 S+E & (1)  &12.9 & 1.46 & 0.49  &0.39 & 35.45 & 3.43 & 3.007 & 3.329 & 2.17 & 1.33 &  -13284.  \\
& (2) &   - &  4.44 & 1.31 & 0.80 & 2.27&  6.83  & 2.786&  3.917  & & &\\  
\\
S+G & (1) & 10.6 & 1.52 & 0.50 & 0.39& 37.09 &  4.23 &   2.524 & 2.740 &2.25 & 1.79&-13817.\\
& (2) &  - &  4.28 & 1.48  &1.46 &1.56 & 6.96  & 3.370 & 2.467 & & & \\    
\\
E+E & (1) &  9.2 & 1.76&  0.32 & 0.24 & 48.51 & 2.61 &   2.915 & 4.077 &  2.36 & 1.38& -14054.\\
& (2) & - &  4.65  &1.37 & 1.01 &2.06 & 6.91 &  4.471&  3.525 &  & &\\
\\
E+S & (1) & 9.1&  1.02  &0.30&  0.22& 64.96&  2.95 &  3.015  &3.130 &2.45 & 0.78 & -14583. \\
  & (2) & - &  4.66 & 1.26 & 1.18 &1.91&  6.95 &  3.093 & 3.403   & & &\\
\\
E+G & (1) & 9.6 & 1.04 & 0.30 & 0.22 & 64.57 & 2.76&  4.333 & 2.385  & 2.43  &0.57 &  -14652. \\
 & (2) & - &  4.24 & 1.10 & 0.89 &1.46 & 6.85 &  3.574 & 3.989 & & &\\       
    
\hline
St. Dev. & (1) & 3.63 & 0.54 & 0.11 & 0.03 & 0.48 & 1.02 & 0.79 & 0.79 & 0.08 & 0.22 & 80.\\
                        & (2) & 0.85 & 0.38 & 0.04 & 0.03 & 0.014 & 0.07 & 0.84 & 0.68 \\

\hline

\end{tabular}
\end{center}
\end{table*}

Among the best combinations obtained for the best model shapes, the parameters are very similar: the best fit is obtained with the two ellipsoids having an orientation of about 9\deg\ to 13\deg, the first ellipsoid with a scale length  x$_{0}$ of 1.6 and secondary scale length of 0.5 and 0.4 kpc. Values are slightly different when an exponential shape is assumed, because the exponential is significantly more like a disc than a 
sech$^{2}$ or a Gaussian, and the different shape  is compensated  for by different scale lengths and normalization. { Regarding the likelihood, the best solution is with a sech$^{2}$ for the first component and an exponential for the second one.}

The second ellipsoid is generally less massive  than the first one but with much larger scale lengths (4./1.3/0.8). 
The disc scale length is about 2.2 to 2.3 kpc with a wide hole of about 1 to 1.2 kpc (but 0.5 kpc only when an E-shape is assumed for the bulge). As a result, the maximum of the disc density appears to be outside the bulge/bar maximum, as if the internal disc was swept out by the bar. The boxiness of the two ellipsoids is noticeable (\cpara and \cperp generally larger than 2) but their values are not very well constrained. 

Maps of the star counts in the K$_{\rm s}$ band  are shown in figure~\ref{2aligned}, to be compared with fig.~\ref{1pop}. Residuals are much smaller than in the 1-ellipsoid model, and the fit appears good in nearly all regions with residuals at the level of 10\%.
 The population of the nuclear bar at $|l|<2$\deg\ as in previous model appears by subtraction. To see where the differences are most significant, accounting for the Poisson noise, we have plotted the difference divided by the Poissonian counting error of the model (i.e. ($N_{\rm obs}-N_{\rm mod})/\sqrt{N_{\rm mod}}$) in the bottom right-hand panel of figure~\ref{2aligned}.

  \begin{figure*}
   \centering
   \includegraphics[width=7cm]{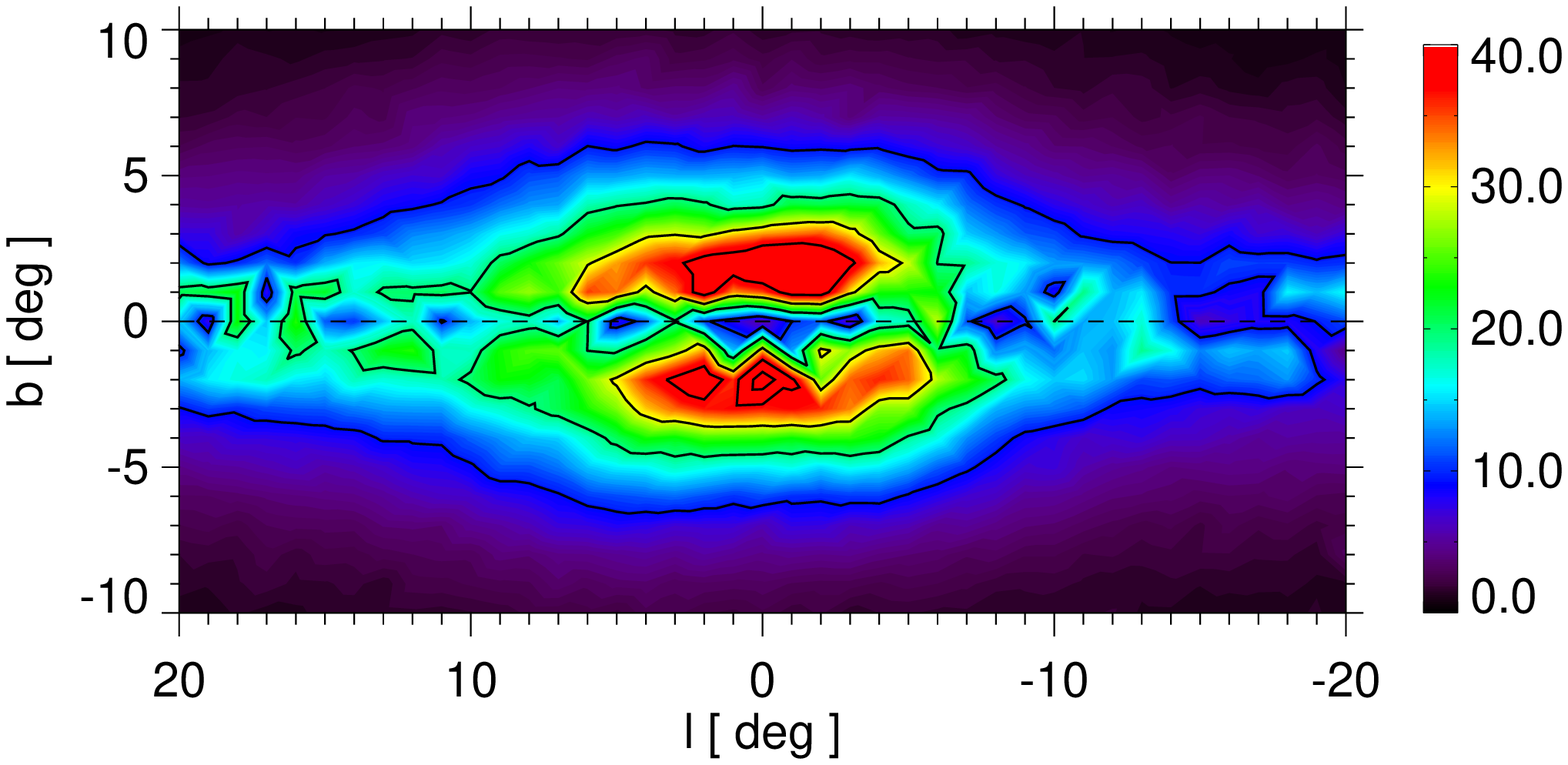}
    \includegraphics[width=7cm]{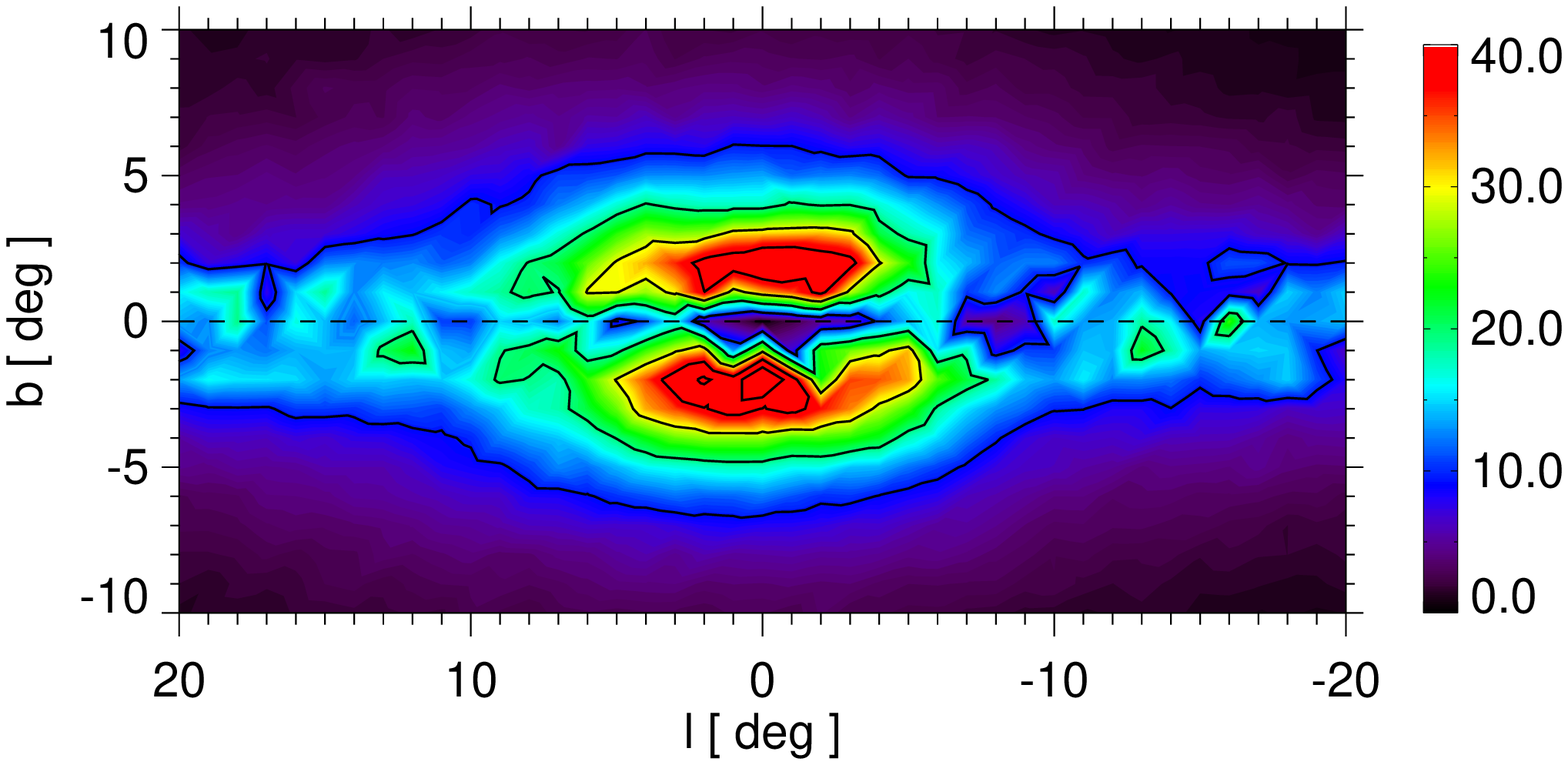}
     \includegraphics[width=7cm]{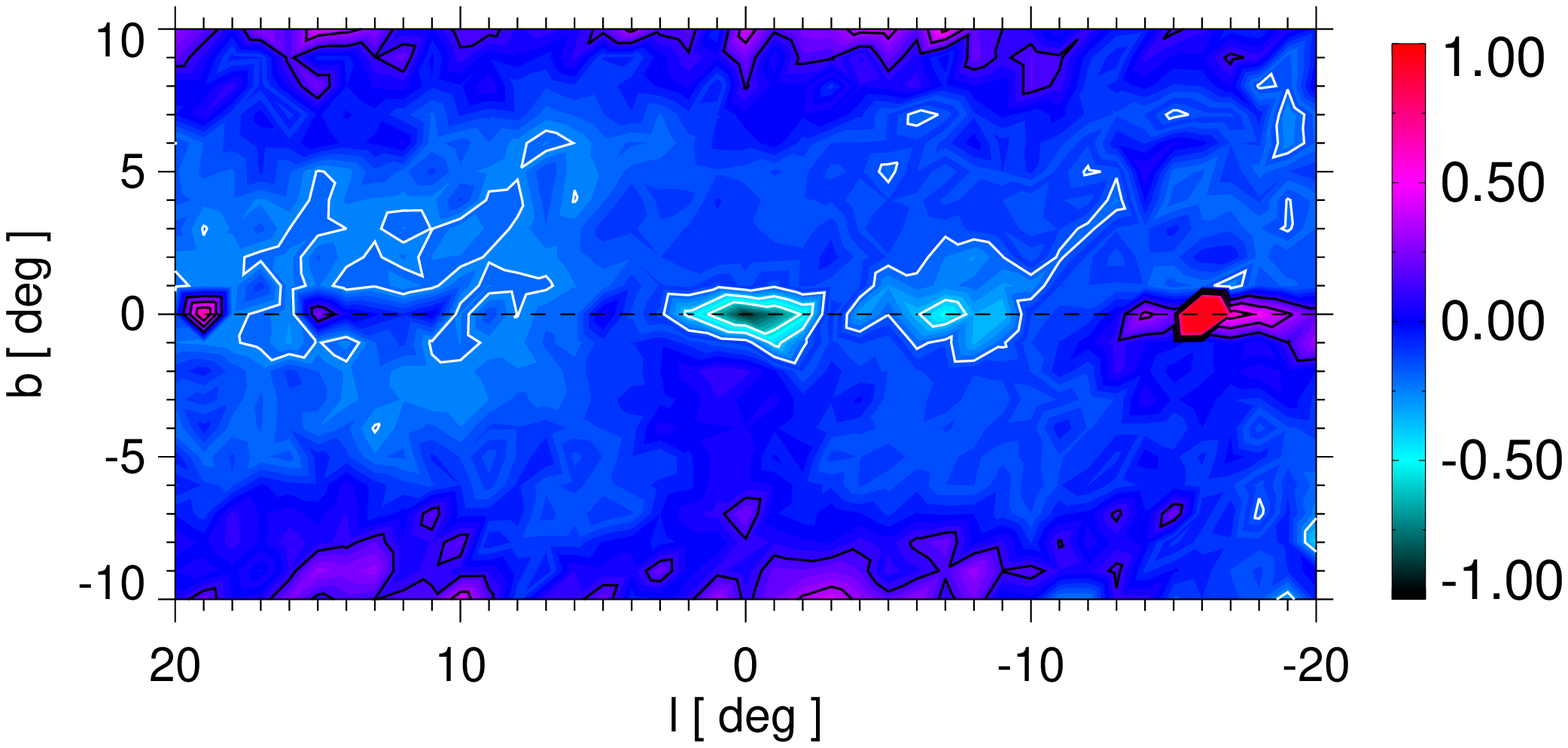}
   \includegraphics[width=7cm]{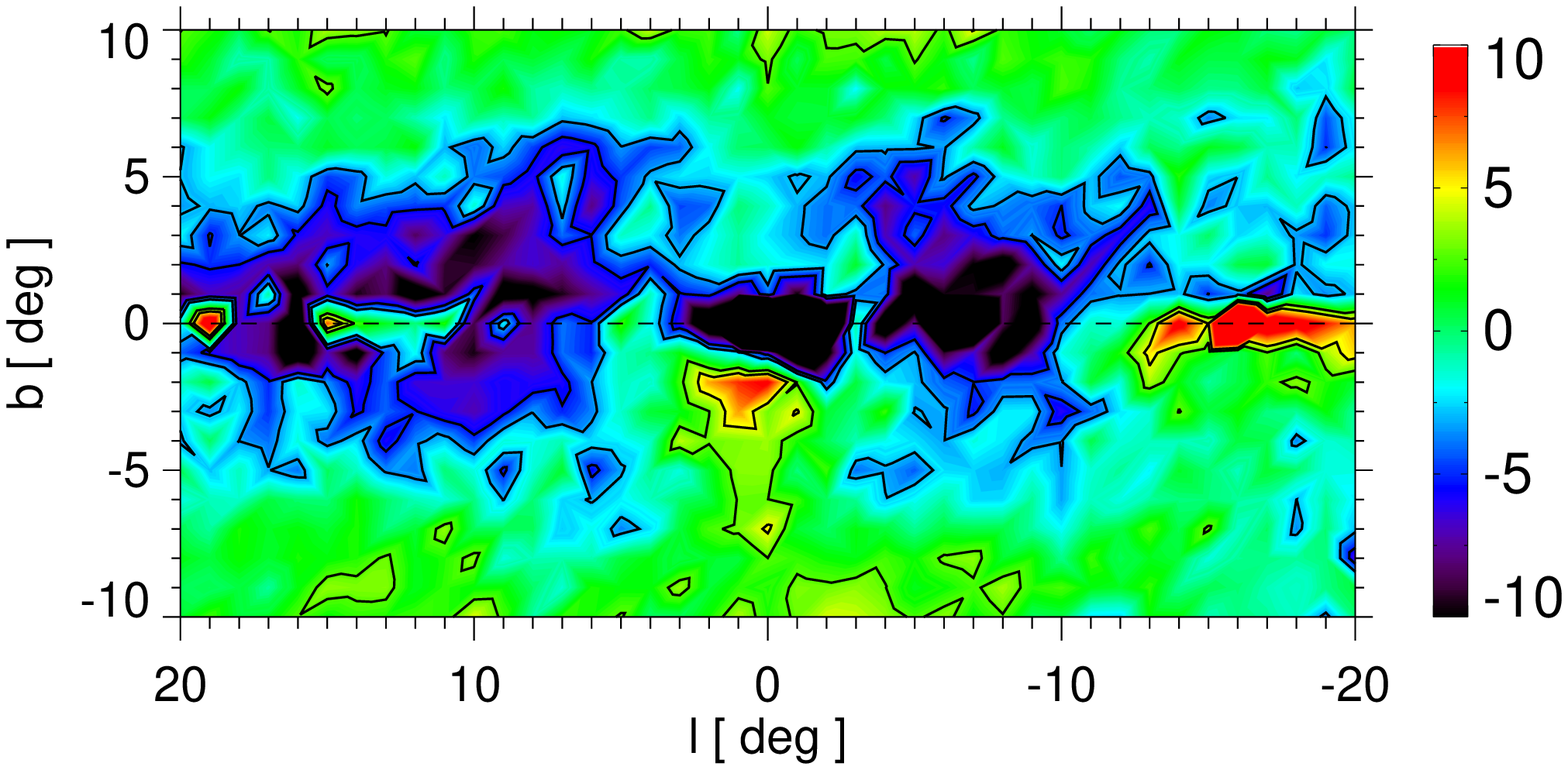}
      \caption{Same as figure~\ref{1pop} but with two fitted ellipsoids.}
       \label{2aligned}
   \end{figure*}

\subsection{Attempts with other shapes}

As is the case for many disc galaxies, the Milky Way has a spiral structure even if it is not well defined at present. The spiral arms are mainly detected  in the visible from young populations and from interstellar matter tracers such as HII regions. 2MASS star counts are dominated by K giants, which generally do not follow the spiral structure because of their age, but the counts could be slightly sensitive to them because in any case, a part of these K giants are indeed young objects. Another structure that has been  well studied in CO is the molecular ring where a large proportion of the giant molecular clouds reside. It is not well established whether the stellar populations contribute to this ring. \cite{Sevenster2001} propose that a stellar ring significantly enhances the microlensing optical depth towards the bulge.

To check this point we investigate the model likelihood when a ring shape is included in the fit in place of the second ellipsoid. For this we use the \cite{Sevenster2001} elliptical ring density law, with free parameters: ring mean radius, ellipticity, and orientation in the plane, thickness in the plane, scale height, and normalization. We also tested cutoffs in longitude for this ring to allow for the existence of a portion of ring rather than a complete one. The result is that the likelihood is comparable to the 1 ellipsoid model, and the density of such a structure is found to be compatible with 0.  It means that in the longitude and latitude zones considered here, the ring and the spiral arms do not significantly contribute to the 2MASS star counts.

\begin{figure*}[ht]


            \includegraphics[bb= 50 50 194 149, angle=0,scale=0.47]{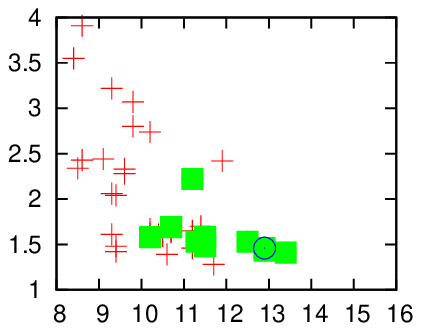}
      \mbox{${x_{0}}$}
     
      \includegraphics[bb= 50 50 194 149, angle=0,scale=0.47]{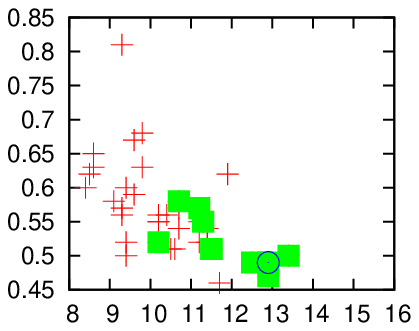}
      \includegraphics[bb= 50 50 194 149, angle=0,scale=0.47]{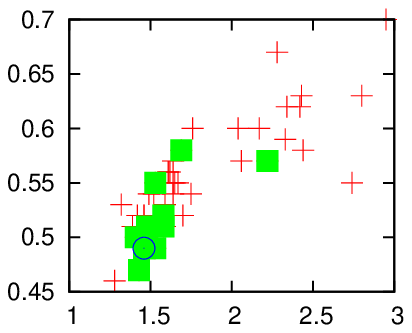}
      \mbox{${y_{0}}$}

            \includegraphics[bb= 50 50 194 149, angle=0,scale=0.47]{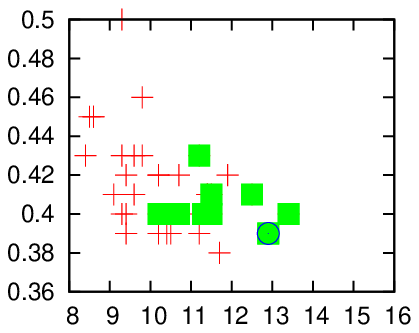}
      \includegraphics[bb= 50 50 194 149, angle=0,scale=0.47]{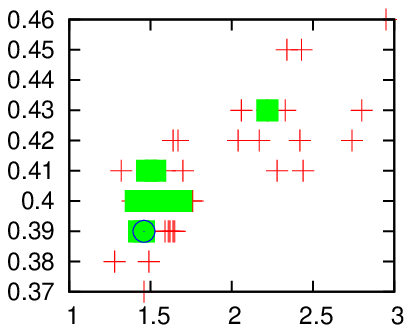}
      \includegraphics[bb= 50 50 194 149, angle=0,scale=0.47]{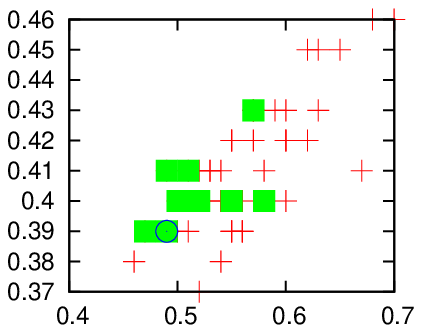}
      \mbox{${z_{0}}$}

      \includegraphics[bb= 50 50 194 149, angle=0,scale=0.47]{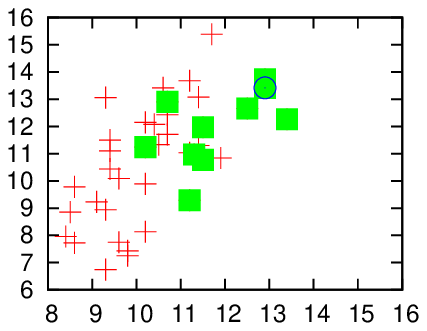}
      \includegraphics[bb= 50 50 194 149, angle=0,scale=0.47]{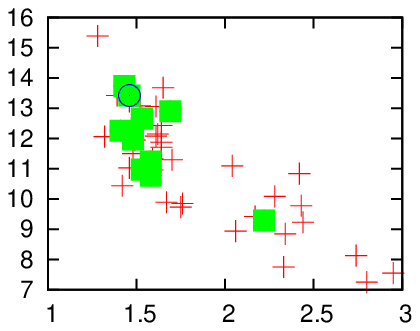}
      \includegraphics[bb= 50 50 194 149, angle=0,scale=0.47]{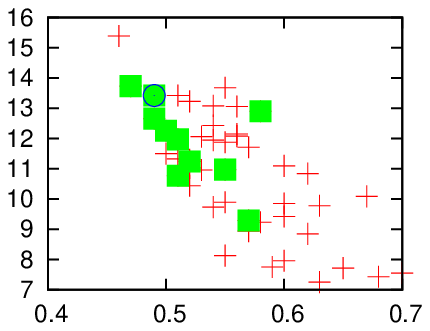}
      \includegraphics[bb= 50 50 194 149, angle=0,scale=0.47]{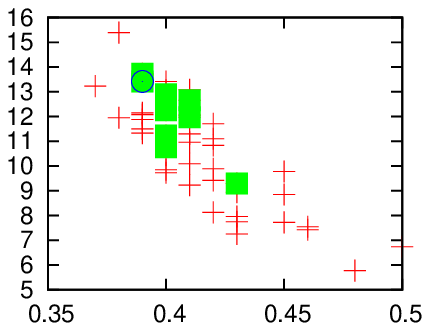}
      \mbox{${{Normalization}}$}

        \includegraphics[bb= 50 50 194 149, angle=0,scale=0.47]{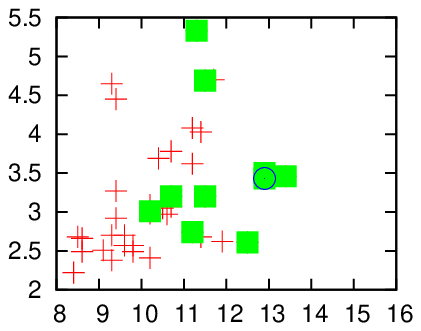}
       \includegraphics[bb= 50 50 194 149, angle=0,scale=0.47]{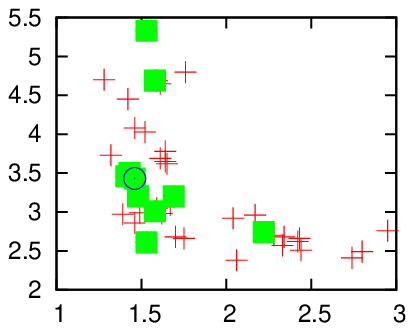}
      \includegraphics[bb= 50 50 194 149, angle=0,scale=0.47]{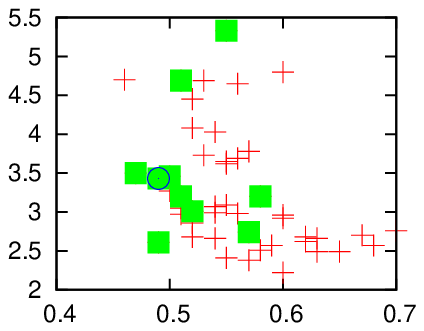}
      \includegraphics[bb= 50 50 194 149, angle=0,scale=0.47]{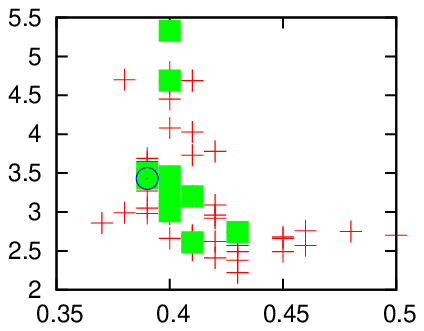}
      \includegraphics[bb= 50 50 194 149, angle=0,scale=0.47]{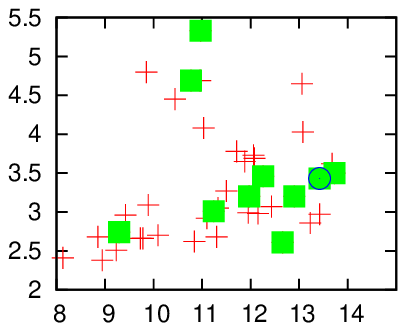}
      \mbox{${Rcut}$}

      \includegraphics[bb= 50 50 194 149, angle=0,scale=0.47]{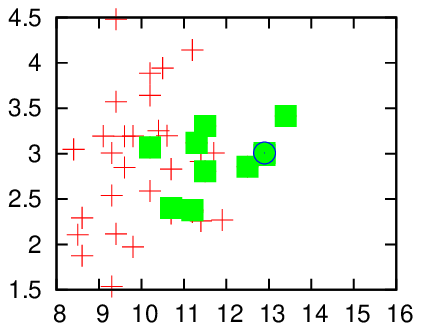}
      \includegraphics[bb= 50 50 194 149, angle=0,scale=0.47]{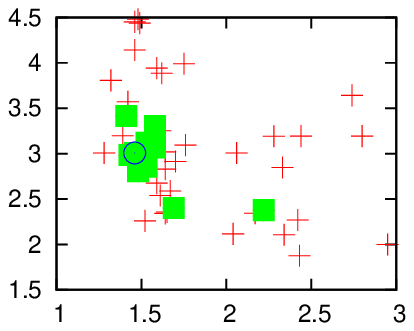}
       \includegraphics[bb= 50 50 194 149, angle=0,scale=0.47]{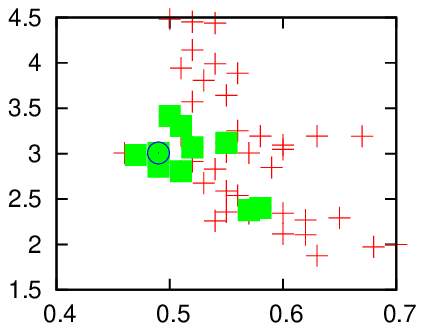}
      \includegraphics[bb= 50 50 194 149, angle=0,scale=0.47]{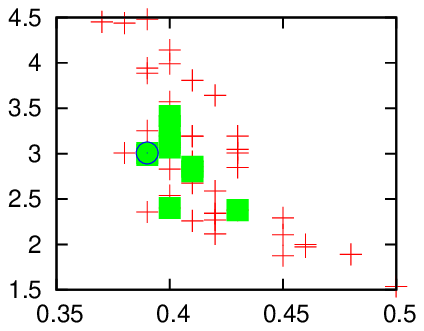}
      \includegraphics[bb= 50 50 194 149, angle=0,scale=0.47]{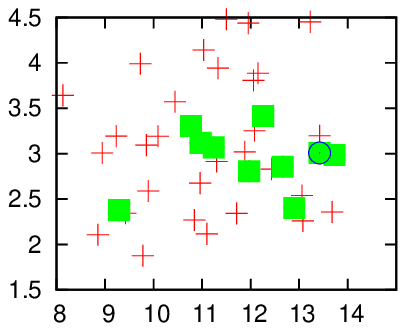}
      \includegraphics[bb= 50 50 194 149, angle=0,scale=0.47]{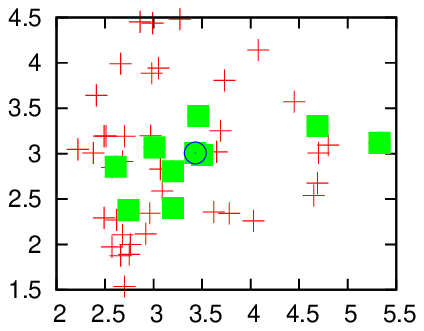}
      \mbox{${c_{\parallel}}$}

       \includegraphics[bb= 50 50 194 149, angle=0,scale=0.47]{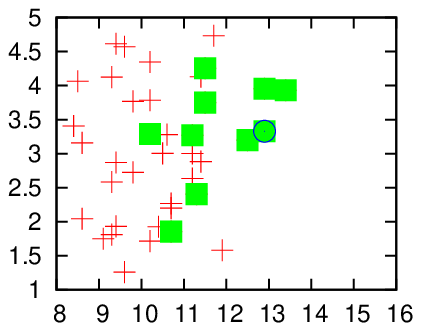}
      \includegraphics[bb= 50 50 194 149, angle=0,scale=0.47]{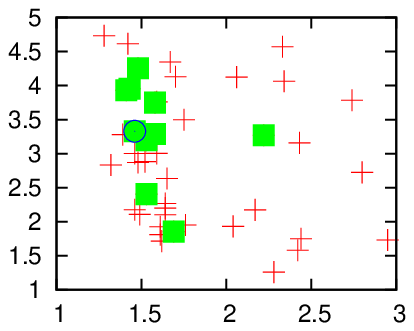}
      \includegraphics[bb= 50 50 194 149, angle=0,scale=0.47]{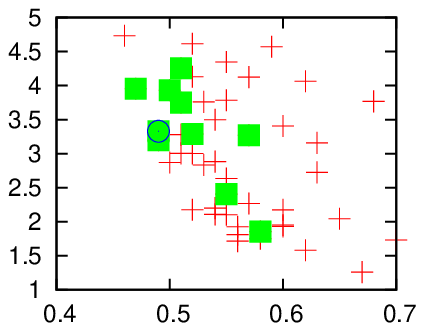}
       \includegraphics[bb= 50 50 194 149, angle=0,scale=0.47]{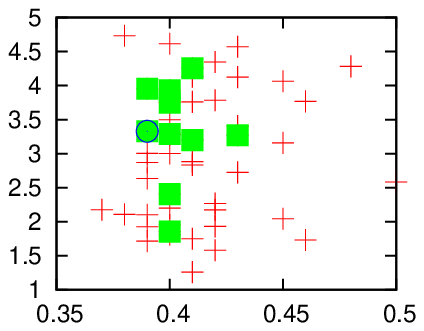}
      \includegraphics[bb= 50 50 194 149, angle=0,scale=0.47]{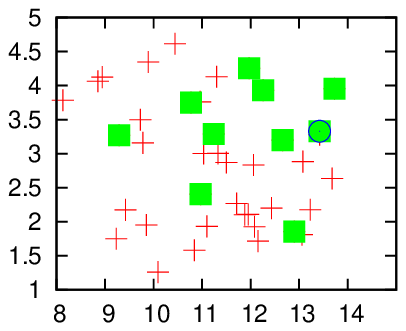}
      \includegraphics[bb= 50 50 194 149, angle=0,scale=0.47]{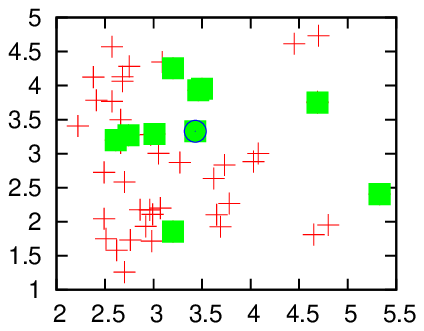}
      \includegraphics[bb= 50 50 194 149, angle=0,scale=0.47]{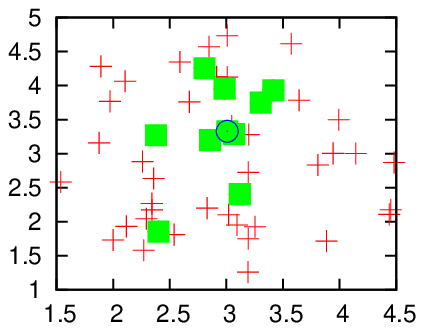}
      \mbox{${c_{\perp}}~$}
     
\mbox{$~~~~~~~~~~~~~~~{\phi}~~~~~~~~~~~~~~~~~~~~~~~~~~~$} \mbox{${x_{0}}~~~~~~~~~~~~~~~~~~~~~~~~~~$} \mbox{${y_{0}}~~~~~~~~~~~~~~~~~~~~~~~~~~~$} \mbox{${z_{0}}~~~~~~~~~~~~~~~~~~~~~~~~$} \mbox{${Norm.}~~~~~~~~~~~~~~~~~~~~~~~~$} \mbox{${Rcut}~~~~~~~~~~~~~~~~~~~$}  \mbox{${c_{\parallel}}~~~~~~$}
    
        \caption{Distribution of solutions for the main bulge population (first ellipsoid) with regards to its parameters. Columns are, from the left to the right: angle in degrees, scale length x0, y0, z0 in kpc, central mass, \cpara and \cperp. Rows are given in the same order from top to bottom. Red plus signs indicate all likelihood, squares hold for the 10 best likelihoods, the blue circle for the best solution. }
       \label{correlation-bulge}
   \end{figure*}
   \begin{figure*}[ht]
\mbox{$^{~\phi}$}

      \includegraphics[bb= 50 50 194 149, angle=0,scale=0.47]{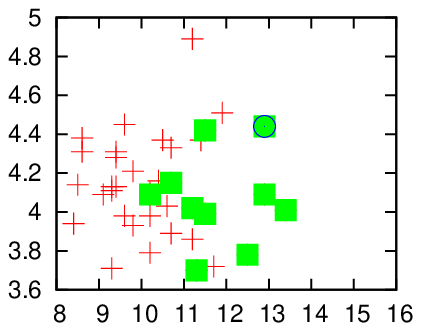}
      \mbox{${{x_{0}}}$}
      
      \includegraphics[bb= 50 50 194 149, angle=0,scale=0.47]{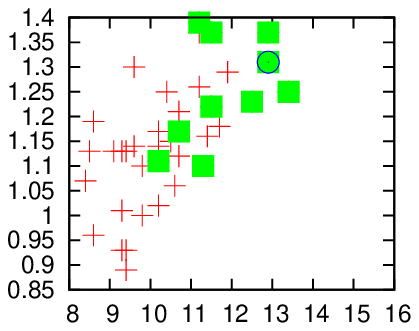}
      \includegraphics[bb= 50 50 194 149, angle=0,scale=0.47]{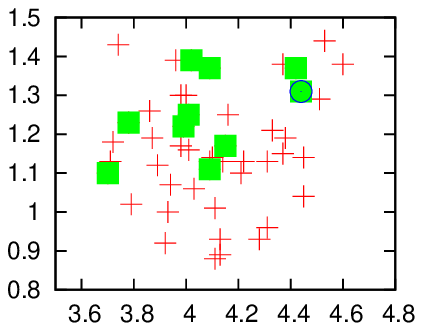}
      \mbox{${y_{0}}$}

      \includegraphics[bb= 50 50 194 149, angle=0,scale=0.47]{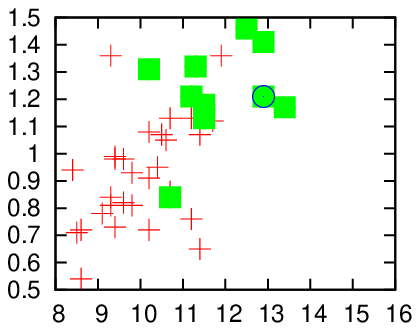}
      \includegraphics[bb= 50 50 194 149, angle=0,scale=0.47]{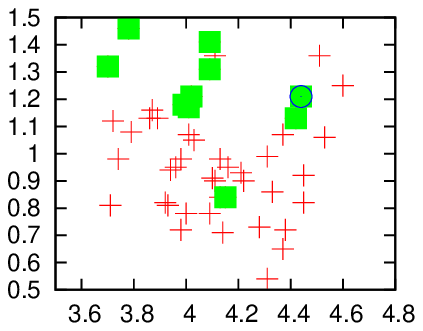}
      \includegraphics[bb= 50 50 194 149, angle=0,scale=0.47]{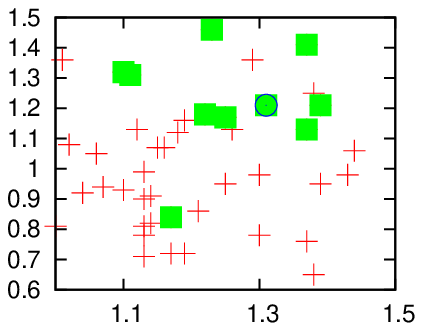}
      \mbox{${z_{0}}$}

      \includegraphics[bb= 50 50 194 149, angle=0,scale=0.47]{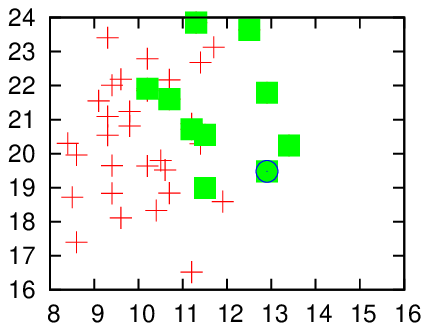}
      \includegraphics[bb= 50 50 194 149, angle=0,scale=0.47]{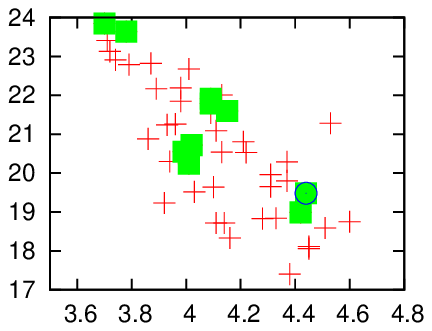}
      \includegraphics[bb= 50 50 194 149, angle=0,scale=0.47]{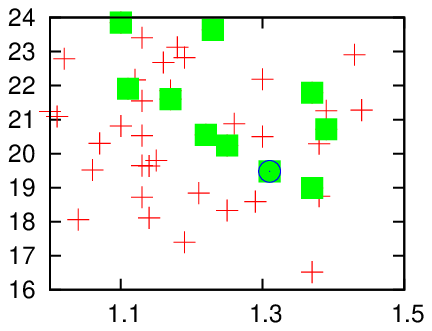}
      \includegraphics[bb= 50 50 194 149, angle=0,scale=0.47]{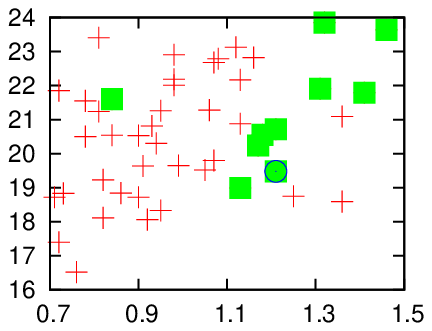}
      \mbox{${Normalization}$}

      \includegraphics[bb= 50 50 194 149, angle=0,scale=0.47]{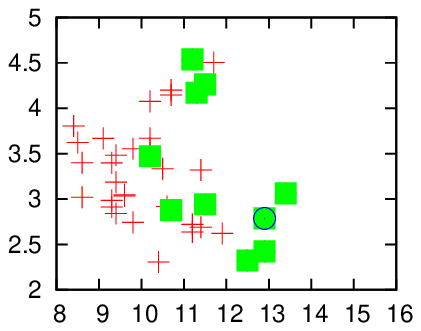}
      \includegraphics[bb= 50 50 194 149, angle=0,scale=0.47]{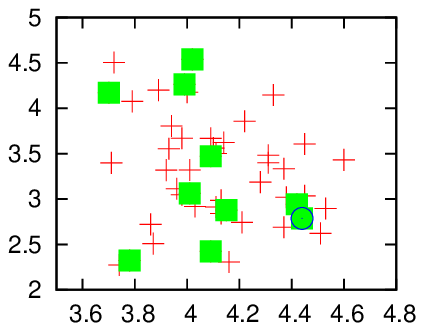}
       \includegraphics[bb= 50 50 194 149, angle=0,scale=0.47]{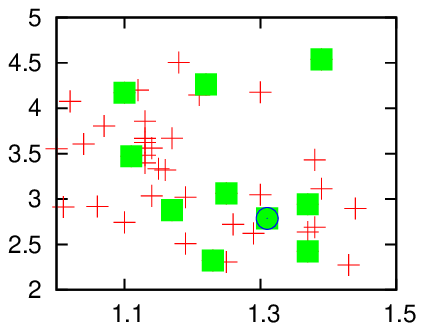}
      \includegraphics[bb= 50 50 194 149, angle=0,scale=0.47]{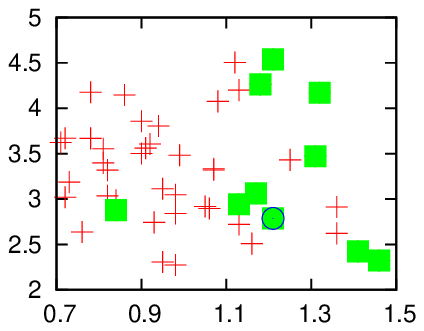}
      \includegraphics[bb= 50 50 194 149, angle=0,scale=0.47]{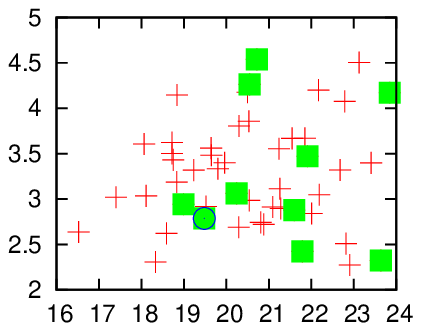}
      \mbox{${c_{\parallel}}$}

       \includegraphics[bb= 50 50 194 149, angle=0,scale=0.47]{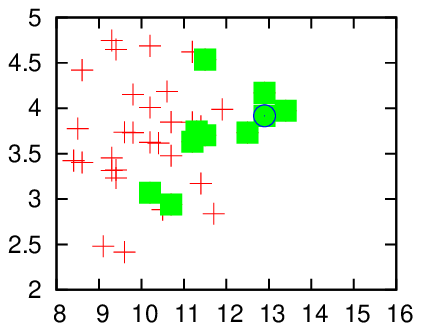}
      \includegraphics[bb= 50 50 194 149, angle=0,scale=0.47]{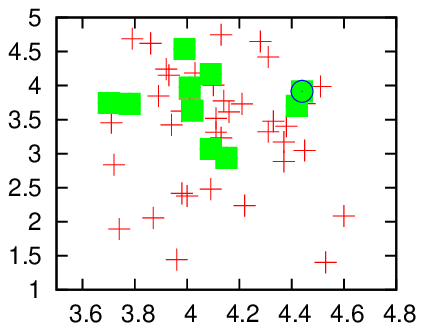}
      \includegraphics[bb= 50 50 194 149, angle=0,scale=0.47]{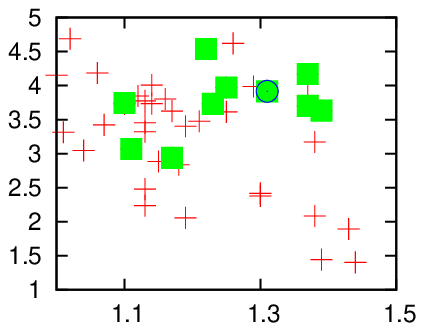}
       \includegraphics[bb= 50 50 194 149, angle=0,scale=0.47]{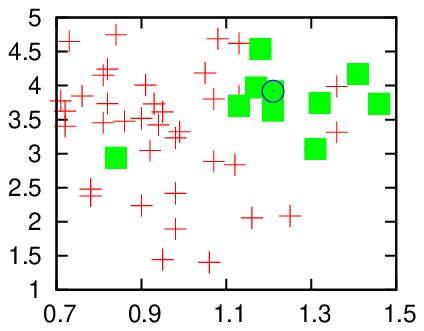}
      \includegraphics[bb= 50 50 194 149, angle=0,scale=0.47]{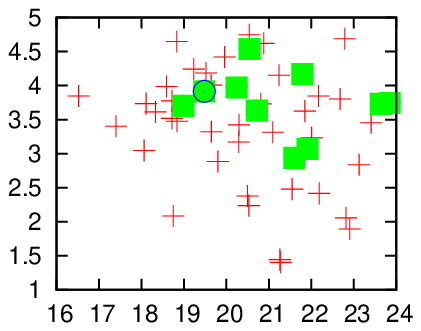}
      \includegraphics[bb= 50 50 194 149, angle=0,scale=0.47]{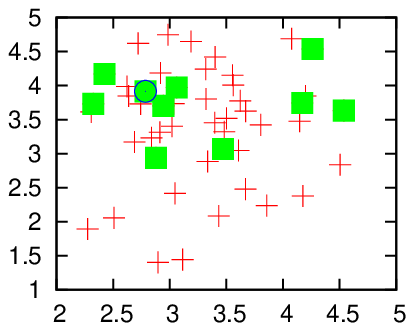}
      \mbox{${c_{\perp}}~$}

\mbox{~~~~~~~~~~~~${\phi}~~~~~~~~~~~~~~~~~~~~~~~~$} \mbox{${x_{0}}~~~~~~~~~~~~~~~~~~~~~~~~~~$} \mbox{${y_{0}}~~~~~~~~~~~~~~~~~~~~~~~~~~~$} \mbox{${z_{0}}~~~~~~~~~~~~~~~~~~~~~~~~~~$} \mbox{${Norm.}~~~~~~~~~~~~~~~~~~~~~~~~~~$}  \mbox{${c_{\parallel}}~~~~~~$}

        \caption{Distribution of solutions for the second ellipsoid with regards to its parameters. Red plus signs indicate all solutions, squares hold for the 10 best likelihood solutions, blue circle for the best solution.}
       \label{correlation-bar}
   \end{figure*}

     \begin{figure*}[ht]

      \includegraphics[bb= 50 50 194 149, angle=0,scale=0.42]{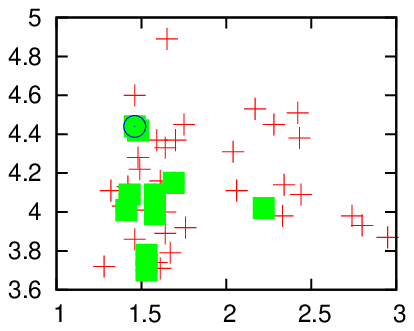}
      \includegraphics[bb= 50 50 194 149, angle=0,scale=0.42]{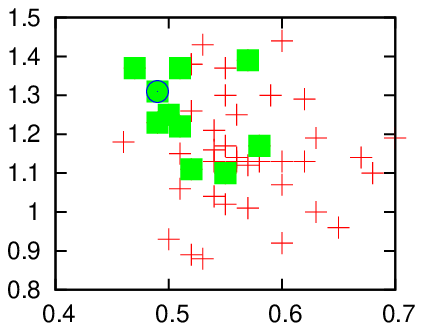}
      \includegraphics[bb= 50 50 194 149, angle=0,scale=0.42]{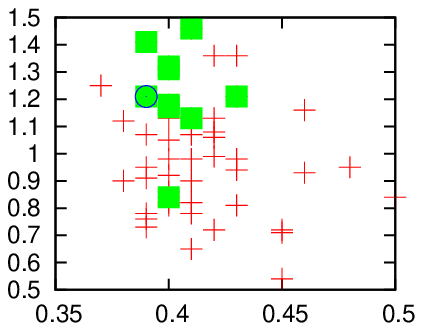}
      \includegraphics[bb= 50 50 194 149, angle=0,scale=0.42]{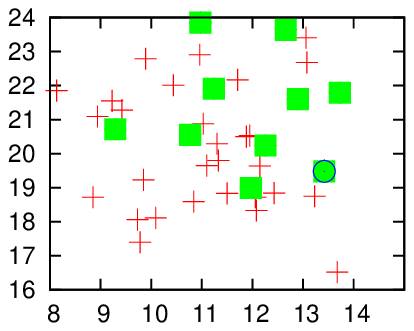}
      \includegraphics[bb= 50 50 194 149, angle=0,scale=0.42]{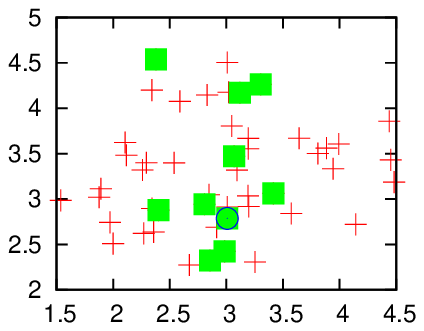}
      \includegraphics[bb= 50 50 194 149, angle=0,scale=0.42]{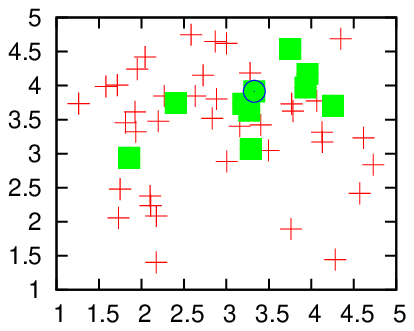}

\mbox{$~~~~~~~~~~~~~~~$} \mbox{${x_{0}}~~~~~~~~~~~~~~~~~~~~~~$} \mbox{${y_{0}}~~~~~~~~~~~~~~~~~~~~~~~$} \mbox{${z_{0}}~~~~~~~~~~~~~~~~~~~~~~$} \mbox{${Norm.}~~~~~~~~~~~~~~~~~~~~$}  \mbox{${c_{\parallel}}~~~~~~~~~~~~~~~~~~~~~$} \mbox{${c_{\perp}}~~~~~~$}

        \caption{Distribution of solutions and correlation between parameters of the second ellipsoid with regard to the first ellipsoid. The first  ellipsoid is in abscissa, second  in ordinate. Red plus signs indicate all likelihoods, green squares hold for the 10 best likelihoods, blue circle for the best solution.  }
       \label{correlation-bulge-bar}
   \end{figure*}

\subsection{Uncertainties and correlations}

Using our method we are able to estimate both the accuracy with which the parameters are determined and the correlations between them. A Monte-Carlo process was computed about 40 times in the case of the best model (at least 20 times in other cases) to compute the average best parameters and give estimates of the uncertainties and correlations between parameters.
For the model S+E, correlations between parameters are shown in figure~\ref{correlation-bulge} for the first ellipsoid and in figure~\ref{correlation-bar} for the second one. Cross-correlations between  parameters of the two ellipsoids are shown in figure~\ref{correlation-bulge-bar}. 
 In these figures for all trials, the best ten solutions (in the sense of the likelihood) and the best likelihood solution shown in the table. The correlations between parameters of a given population are rather weak in general, indicating that the method is robust. The only net correlation appears between scale lengths for the first ellipsoid (between x$_0$ and y$_0$ and z$_0$) and with the normalization.

The uncertainties on the scale lengths is about 250 pc for the major axis, and only 50 pc on the secondary axis. The normalization is determined at the level of 15\% for the main ellipsoid and the second ellipsoid.
The boxyness parameters \cpara and \cperp are nearly always found to be larger than two, being three in the mean, making both ellipsoids rather boxy, even if the secondary one is found with an exponential shape (which is more like a disc  than the sech$^{2}$).

The disc and hole-scale lengths are slightly anti-correlated, a  longer scale length favouring a smaller hole. This is expected due to the uncertainty on the distance indicators in broad band photometry. A better fit of the disc scale length would need to include larger longitudes. But then the spiral structure should also be included, as we would reach arm tangents. The model also has a slight tendancy to produce too many stars at $l<-5$\deg\ and too few stars at $l>5$\deg\ close to the plane. We could interpret this by the ability of  the hole to be ellipsoidal in the plane, as found by \cite{Binney97}. We attempted to model this ellipticity but it is poorly constrained, and the fit is only very slightly better than with a round hole. This case could be considered later in more detail with deeper data.

\section{Extinction validation}

\begin{figure}[]
   \centering
   \includegraphics[width=7cm]{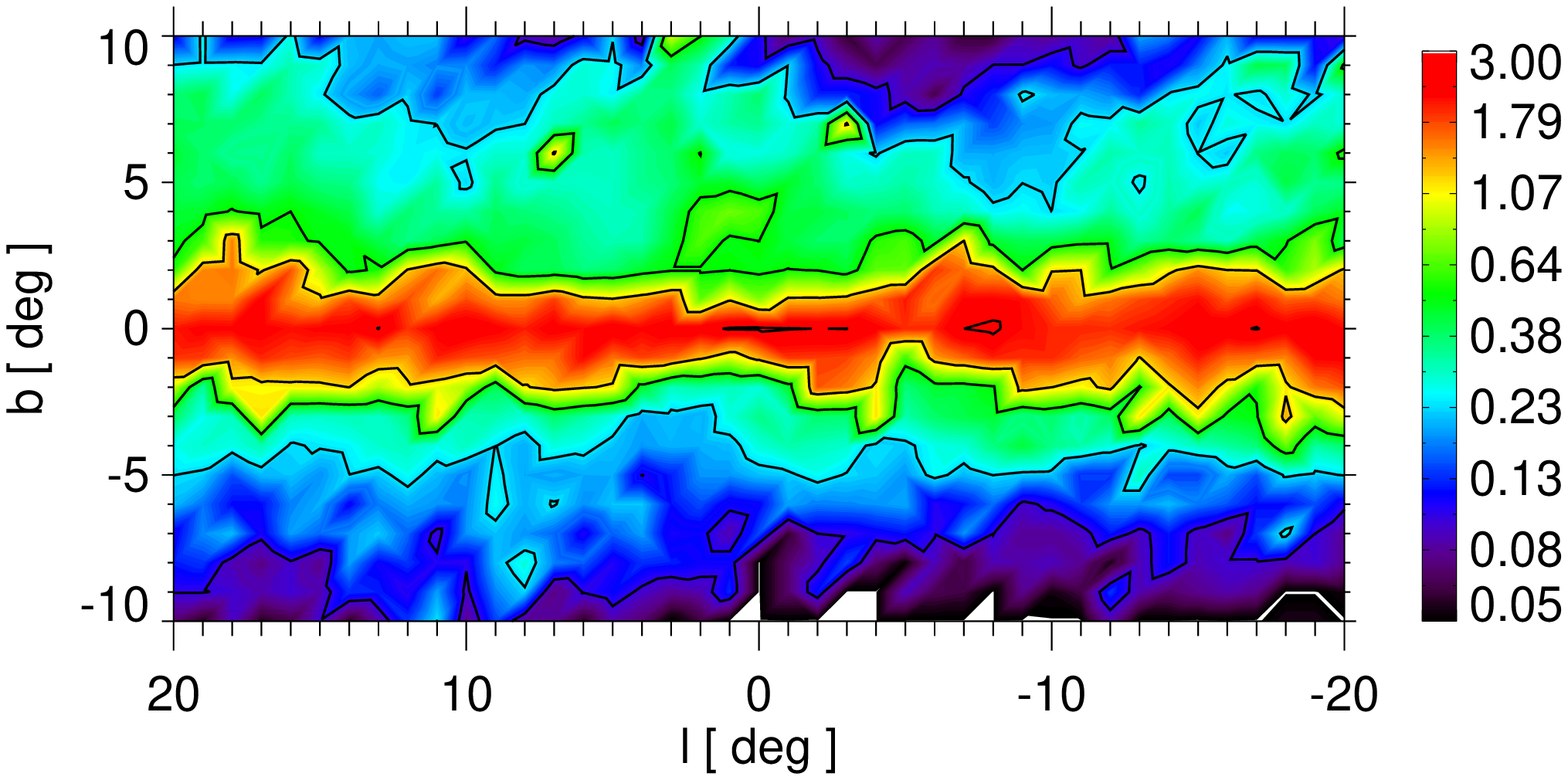}
      \includegraphics[width=7cm]{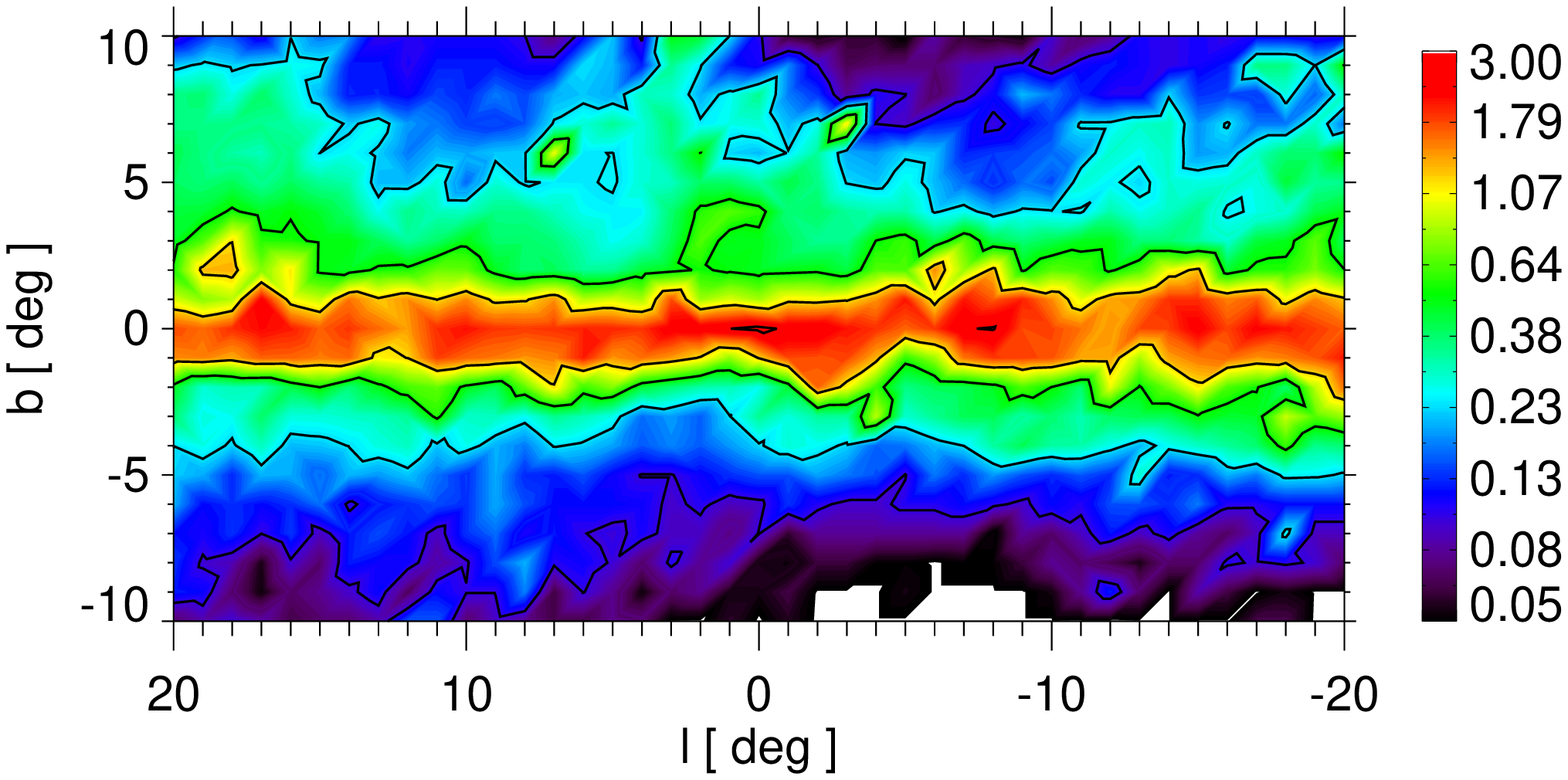}

      \caption{A$_{\rm k}$ map deduced from Marshall et al. (2006) method. Top: based on the standard stellar population model from BGM; Bottom: based on the new stellar population model.}
       \label{akmap}
   \end{figure}

\begin{figure}[]
   \centering
   \includegraphics[width=9cm]{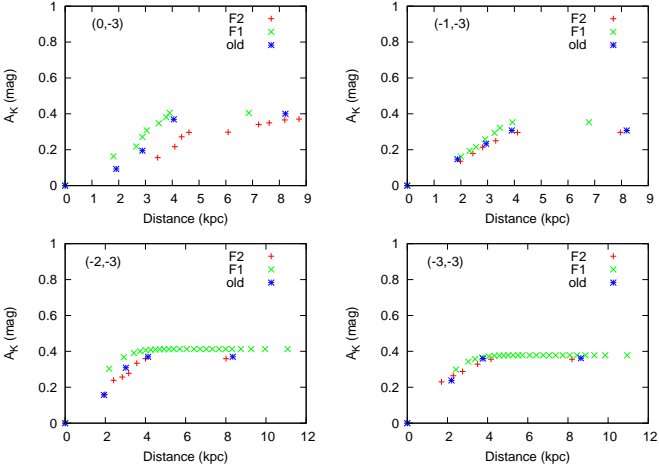}      

      \caption{Distribution of extinction over four lines of sight (latitudes b=-3\deg\  and longitudes between -3 and 0\deg) from the orignal model (Old), 1-ellipsoid model (F1), 2-ellipsoid model (F2). }
       \label{ext-distance}
   \end{figure}

The analysis was done assuming a 3D extinction distribution model obtained using the bulge model from \cite{Picaud04}. We need to ensure that the extinction resulting from the new fit is consistent with the initial extinction we assumed. Figure~\ref{akmap} shows the comparison between the projected extinction map used originally with the map deduced from the fit with the new model. The differences are negligible. We also compared the distribution in distance of A$_{K}$ along some lines of sight and saw generally good agreement in the distances of the clouds, within the error bars of this distance (a few hundred parsecs). Figure~\ref{ext-distance} shows the variation in the distribution of extinction along four lines of sight for the standard model (old), the model with the bulge population simulated by one ellipsoid (F1), and the model with two ellipsoids (model~2, F2). The differences are generally negligible. In one case (longitude 0), we see that the cloud is found at a greater distance with the new model, but the total extinction in the bulge remains the same,
so  the extinction deduced using our method \citep{Marshall06} is sufficiently robust and not particularly model dependent. Conversely, we can emphasize that the 3D extinction model is reliable enough to avoid biasing our results concerning the stellar populations study.

   \section{Resulting colour-magnitude diagrams and double-clump feature}
   
   We present colour-magnitude diagrams for  five directions in order to see the subtle differences between 1-ellipsoid and 2-ellipsoid models. 
   To reproduce the natural width of the sequences in the CMD, we have applied a star-by-star variation of the extinction, about the mean extinction at the star distance given by the extinction map. The cosmic variation in the extinction is assumed to be 20\% of the extinction, a value that is estimated by trial and error.
      
\begin{figure*}
\vskip 1 cm
\includegraphics[]{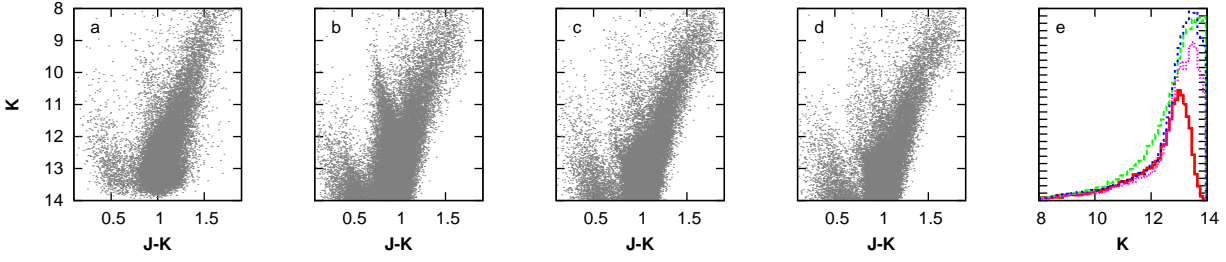}

\caption{Colour-magnitude diagrams for the 2MASS field at l=0, b=-4. (a) data; (b) best-fit model with 1 ellipsoid; (c) best-fit models with 2 ellipsoids; (d) modified model with flared bar; (e) histograms of data (red solid) and models (1 ellipsoid: green long dashed; 2 ellipsoids: blue dotted, flared bar: magenta short dashed). }

\label{cmd-3mod_p0_m4}
\end{figure*}

\begin{figure*}
\includegraphics[]{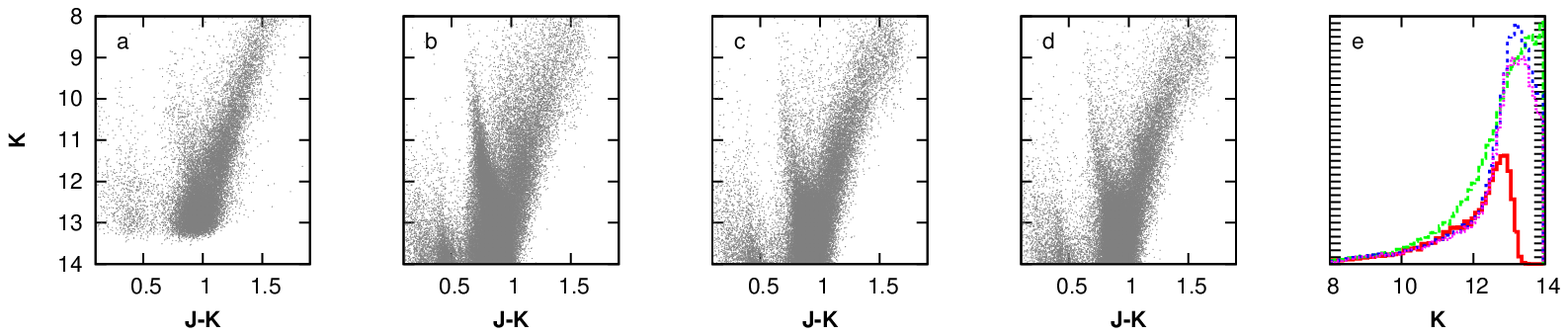}
\vskip 1 cm
\caption{Colour-magnitude diagrams for the 2MASS field at l=+3, b=-3. Same coding as in fig~\ref{cmd-3mod_p0_m4}.}

\label{cmd-3mod_p3_m3}
\end{figure*}

\begin{figure*}
\includegraphics[]{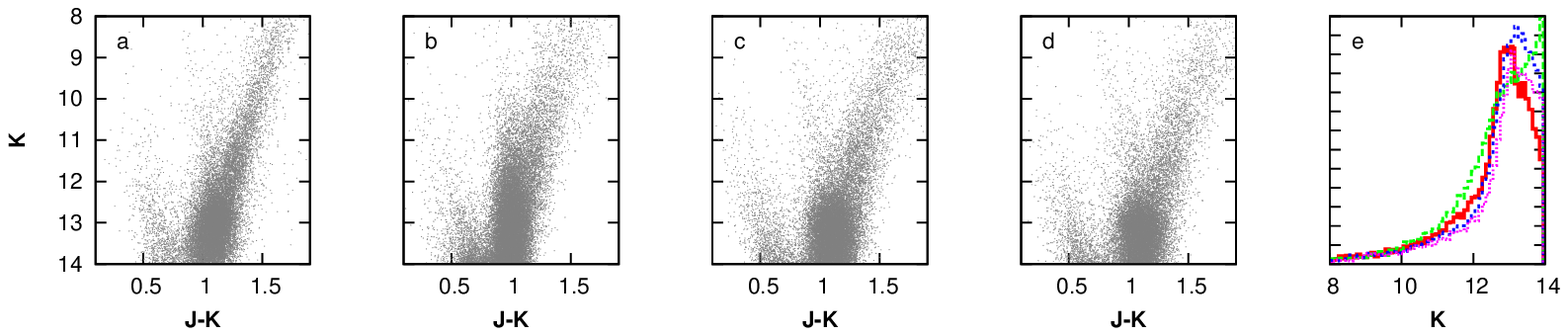}
\vskip 1 cm
\caption{Colour-magnitude diagrams for the 2MASS field at l=5, b=4.5. Same coding as in fig~\ref{cmd-3mod_p0_m4}.}
\label{cmd-3mod_p5_p4.5}
\end{figure*}

\begin{figure*}
\includegraphics[]{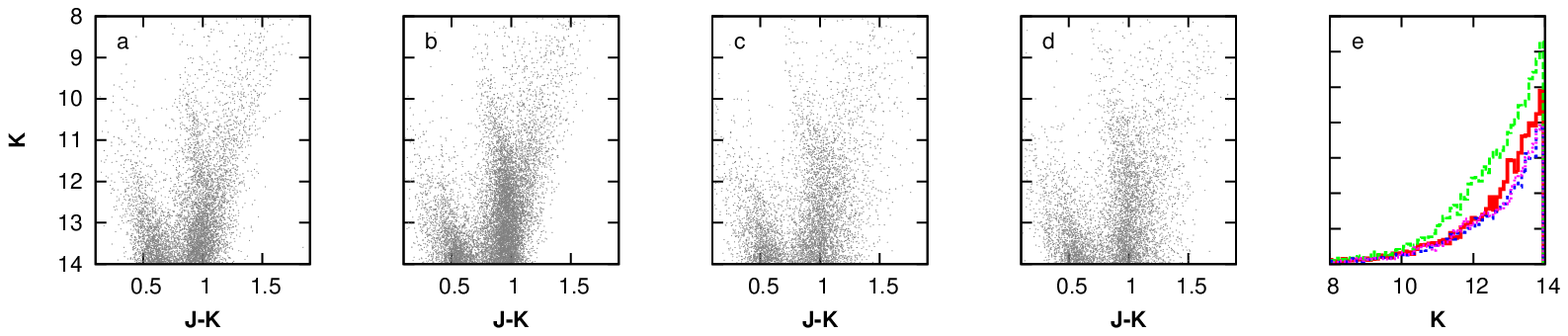}
\vskip 1 cm
\caption{Colour-magnitude diagrams for the 2MASS field at l=-17, b=-4.25. Same coding as in fig~\ref{cmd-3mod_p0_m4}.}
\label{cmd-3mod_m17_m4.25}
\end{figure*}

\begin{figure*}
\includegraphics[]{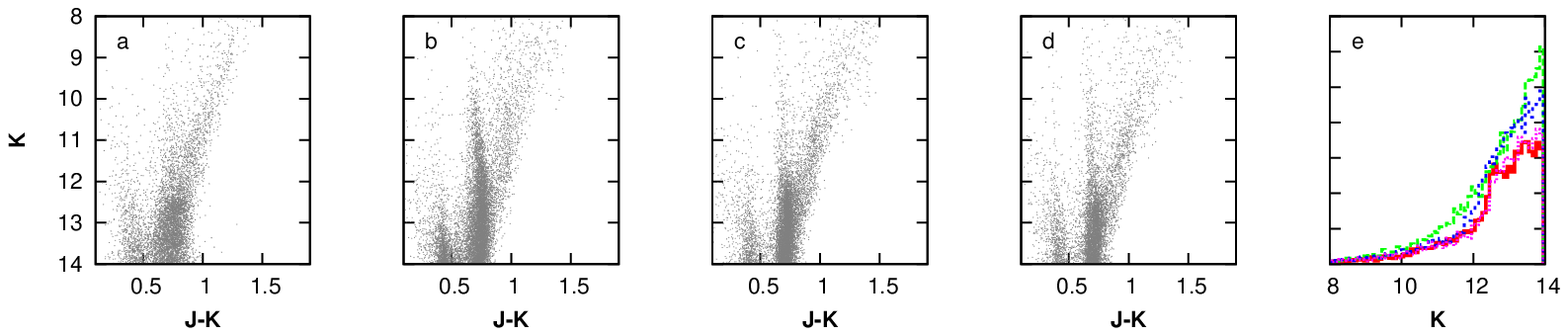}
\vskip 1 cm
\caption{Colour-magnitude diagrams for the 2MASS field at l=0, b=-8. Same coding as in fig~\ref{cmd-3mod_p0_m4}. In this field we see clearly in the histogram the double clump feature reproduced by the modified model (panel d and magenta line) and not by the originally fitted model (panel c and blue line).}
\label{cmd-3mod_p0_m8}
\end{figure*}

   Figures~\ref{cmd-3mod_p0_m4},
   \ref{cmd-3mod_p3_m3},\ref{cmd-3mod_p5_p4.5}, \ref{cmd-3mod_m17_m4.25}, and \ref{cmd-3mod_p0_m8}  show these CMDs. The 2MASS data are in the first columns, 1-ellipsoid model in column 2, 2-ellipsoid model in columns 3, and histograms in K$_{\rm s}$ in column 5. { In column~4 we also present a modified model that  produces a double-clump feature (see below).} One notices the incompleteness of the data starting probably at about K=12.5 in fields at $|b|\le -4$\deg.
   
   The 1-ellipsoid model gives star counts in the red giant branch that are too high.The agreement is generally much better with the 2-ellipsoid model, even if it is not perfect in some fields. There are still residuals in the shape of the giant branch that could probably be solved by varying the star by star dispersion in the extinction, especially in the case of the field with coordinates (l,b) = (5; 4.5). The important point here is to have a general model reproducing most fields, at the expense of some discrepancy in a limited number of fields, since we are interested in the large-scale structure.

\begin{figure}
\includegraphics[width=9cm]{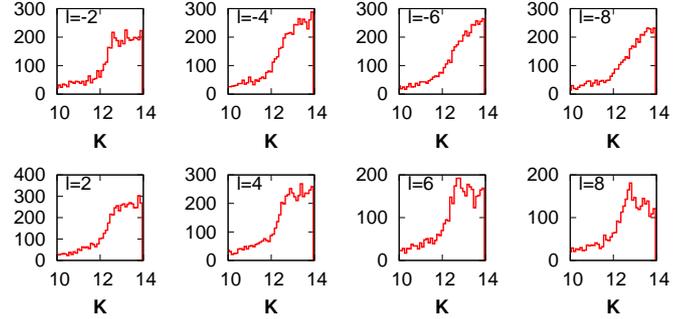}
\caption{Luminosity functions for the red clump region in fields at latitude -8\deg. Coordinates in longitude and latitude are indicated in each panel. The double-clump feature is simulated by adding a slight flare to the original bar model (see text). }
\label{2clump-m8}
\end{figure}

\begin{figure}
\includegraphics[width=7cm]{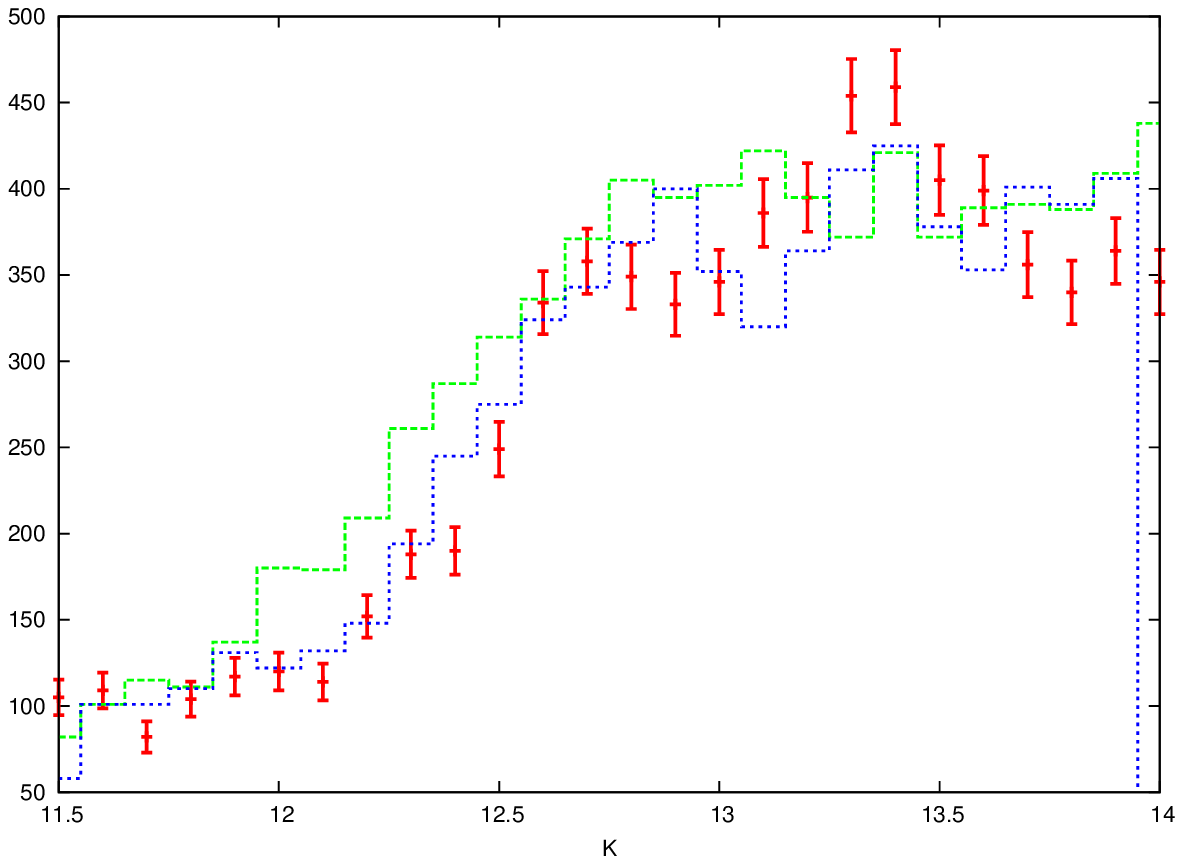}
\caption{Luminosity functions for the red clump region in a field at latitude -7\deg and longitude 0\deg\ from 2MASS (red dots with error bars) compared with the model without flare (in green solid line) and with the flaring bar (in blue dotted line). }
\label{flare-m7}
\end{figure}

{

In fields at intermediate latitudes ($|b|>5$\deg) several studies have shown there are  double clumps. These double clumps change  as a function of longitude \citep{Mcwilliam10}. From our analysis limited to 2MASS differential star counts, and due to the lack of completeness of the data, our solution does not present double clumps, but just a modulation of the shape of the luminosity function due to the shape of the bar. The second structure has its clump peak at fainter magnitudes than the 2MASS completeness limit. If there is a structure creating the double clump it should be in the main bar. 

Since our model reproduces the counts at the level of 10\% in the whole bulge region, we can consider that what is creating the double clump feature is a modulation of the whole shape of the bar. \cite{Pfenniger} did a numerical study of instability in the 3D family of periodic orbits in a barred galaxy model. He shows a family of resonance  creating the growth of a box-shaped bulge. It shows how stars are diffused at high z.  It is natural to hypothesize that the double clump can be produced by such an effect. 

To test this hypothesis, we simulated a flaring bar by enhancing the scale height of the bar with a sinus fonction of the distance along the major axis of the bar. If $x$ is the galactocentric distance along the major axis, and  $dz0$ is the scale height along the $z$ axis, the new vertical scale height is simulated by

\[ dz = dz0\times(1+0.3\times sin(x/650.). \]

\noindent
The original value of $dz0$ (obtained by the fit above) was decreased slightly (from 390 to 330 pc) to account for the fact that we measured the mean scale height when we fixed this parameter in the fit. The parameters in the equation are a reasonable guess in order to fit the clump data at a medium latitude. 

In figure~\ref{2clump-m8} we show the star counts around the clump in various fields at latitude b=-8\deg\  in order to compare with figure~3 of \cite{Mcwilliam10}. It shows that a flaring bar as modelled here can qualitatively reproduce the double-clump feature. The proportion of stars in each clump varies as a function of the longitude due to the flare, and the significant differences between positive and negative longitudes. This figure can be compared with figure 3 of McWilliam \& Zoccali (2010).
It can also be seen on a zoom of the luminosity function in the  field at l=0, b=-7\deg (figure~\ref{flare-m7}) where we superimposed the observed counts with the model predictions without the flare and with the flare. We show that a simple modulation  of our primary model with a slight bar flare reproduces the two-clump feature.

Since the 2MASS data are not complete, we cannot perform a quantitative fit of the flare parameters  with these data. We only show here that a flaring bar can be a realistic explanation for the double-clump  and that it is dynamically reasonable,  as in the \cite{Pfenniger} simulations. In a future paper we plan a more complete analysis of the shape of the flare using VVV data, for example, by comparing simulations with the density plots from \cite{Saito11} in order to specify the characteristics of this flare more accurately  as a function of the galacto-centric distance.
}

\section{The inner bulge}

\begin{figure}[]
   \centering
   \includegraphics[width=9cm]{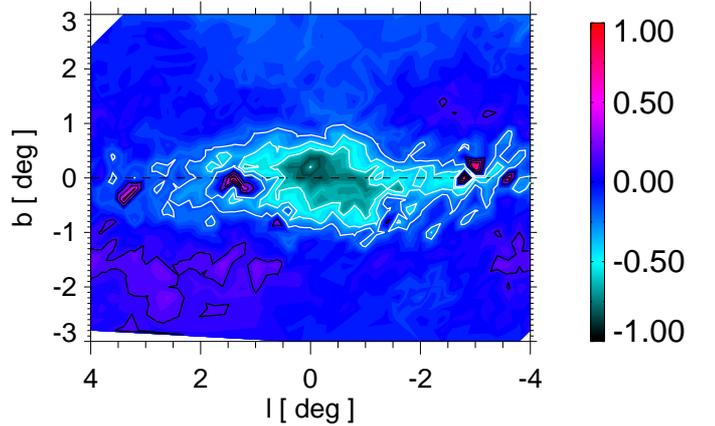}      

      \caption{Difference map for the inner bulge region (N$_{\rm obs}$-N$_{\rm mod}$)/N$_{\rm obs}$. The population in excess in the data is most likely the nuclear bar. }
       \label{nuclear}
   \end{figure}

Assuming that the model of stellar populations and 3D extinction now reproduces 
the populations of the outer bulge and the disc, we can investigate if any further structures are missing from our modelling. As a check we plot in figure~\ref{nuclear} the higher resolution map of the difference between model predictions and observed star counts in the region -4\deg$<$l$<$4\deg\ and -3\deg$<$b$<$3\deg. The excess of stars in the data should represent the population of the inner bulge. It resembles the population seen by \cite{Alard} and coincide with the position of the central molecular zone. Several authors \citep{Morris96,An2011} have shown that this region contains young stellar populations. To investigate further this population, one would need to improve the 3D extinction model here. As we saw in figure~\ref{akmap}, the extinction determined by our method saturates at about A$_{K}$=3 due to the limiting magnitude in 2MASS. Thus our extinction is most probably underestimated, implying that the density of the inner bulge population is overestimated. We do not investigate further the density of this population in this paper, as it could only be qualitative from 2MASS data alone.
We envisage to apply our method to deeper data and at larger wavelengths, like Spitzer at 3.6 and 4.5 microns, in order to make a deeper 3D map of the extinction in this region and to study the inner bulge region, the probable nuclear bar or ring and its link with the central molecular zone.

\section{Discussion}

{ We have shown in this analysis that the bulge region can be simulated using the sum of two ellipsoids: a main component that is a quite standard boxy bulge/bar, which dominates the counts up to latitudes of about 5\deg, and a second ellipsoid that is a thicker structure seen mainly at higher latitudes.} This thick component seems to be about as thick as the local thick disc. 
It could be either the classical bulge, flattened by the effect of the barred potential, or the inner counter part of the thick disc population. To distinguish how these structures can be differentiated and to understand which population they belong to, it is useful to look at the spectroscopic surveys showing metallicity distributions at different latitudes.

\subsection{Metallicity as a function of latitude}

Among studies measuring the metallicity distribution function (MDF) in different fields of the bulge region several have considered the characteristics of the populations along the minor axis at different latitudes. \cite{Zoccali2008} studied the MDF and radial velocity distribution from the RGB, \cite{Gonzales2011} completed the data with the abundances of alpha abundances, 
\cite{Babusiaux2010} studied the correlation between metallicity and kinematics and \cite{Hill2011} computed the MDF from the red clump giants for Baade's window. The overall conclusion of these studies is that  the MDF looks bimodal and the contribution of the components changes with distance from the plane. At a low latitude (-4\deg), the dominant population has a metallicity slightly above-solar and a small contribution comes from a second component of metallicity around -0.5. \cite{Babusiaux2010} show that the high-metallicity component has a high vertex deviation compared to the low-metallicity component that does not. They conclude that the vertical metallicity gradient is the  effect of mixing of populations. On the other hand, \cite{Ness} decompose the bulge populations from an analysis of the ARGOS survey in up to five  components to reproduce the complete MDF. They find a smaller gradient ( about 0.4 dex/kpc) compared with \cite{Zoccali2008}, who found a mean gradient of 0.6 dex/kpc. 

Can we explain these MDFs by our two-ellipsoid model? To answer this question, we simulated the observed MDF from \cite{Zoccali2008} in three latitude ranges, on the minor axis and deduced what would be the necessary metallicity of each of our components to explain the observations. 

\begin{figure*}[]
   \centering
   \includegraphics[width=15cm]{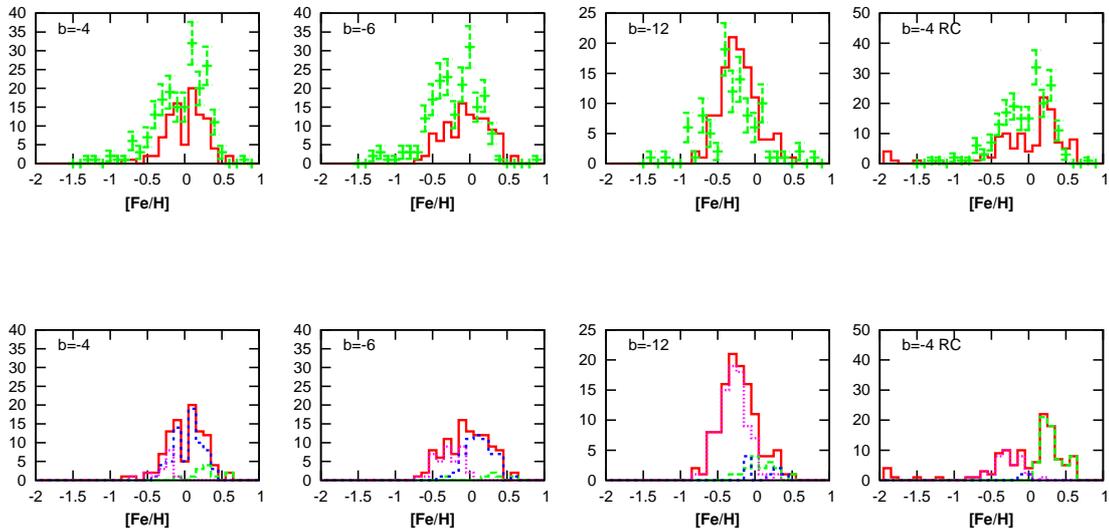}      

      \caption{Metallicity distribution for RGB samples from Zoccali et al (2008) in fields at l=0\deg\ and b=-4\deg, -6\deg, and -12\deg, and for the red clump samples from Hill et al. (2011) in Baade's window (columns from left to right). {\it Top}: The observed data (green dots with Poisson noise) are compared with model simulations, applying the same selection function as in the observations (red solid lines). {\it Bottom}: decomposition in populations of the model simulation. Red solid line: total; green long dash: disc, short dash blue: boxy bar; dotted magenta: thick bulge. The error bars take only the Poisson noise  into account  and not the observational errors or the metallicity or the error in the selection function estimation.}
       \label{MDF2}
   \end{figure*}

We assume that the boxy bar has a mean solar metallicity and that the thick component has a mean metallicity of -0.35 dex. We did  not fit the density of each population, which is the one obtained from the 2MASS star-count fitting. We applied the selection function explained in \cite{Zoccali2008} for the RGB and in \cite{Hill2011} for the red clump sample.
Figure~\ref{MDF2} shows the comparison between the observed metallicity distribution and the simulated one for RGB giants at l=0\deg and b=-4\deg, -6\deg, and -12\deg, and for red clump giants at b=-4\deg. At the bottom we also show the decomposition of the distribution in the three major components: the thin disc, the ``boxy-bar'', and the ``thick-bulge''. We see  good agreement  beween our model predictions and the data in all four cases, and even clearer at high latitudes, within the small Poisson statistics of the samples. 
{ The error bars are only the Poisson noise on the counts in the selected sample. It does not take  observational errors on the metallicity into account  or the noise due to the selection function}. The only significant disagreement comes from  a lower number of low-metallicity stars in the model for the field at b=-6\deg compared to the data. 
This feature will be explored in more detail as soon as a larger sample will be available. It will then be possible to determine whether a metallicity gradient should be added to one of the two populations  or whether the relative density of the two populations is represented well  by this model or should be revised given the errors. 

\subsection{Conclusion and future plans}

{ We have presented here an attempt to fit differential  star counts, CMDs, and metallicity distribution functions in a wide area over the Milky Way bulge region. The model accounts for the metallicity distribution on the minor axis of the bulge, the dissymmetry of the star counts as a function of longitude, and the existence of double clumps at medium latitudes}. This model has two components. The main one is modelled by boxy, S-shape triaxial Ferrer ellipsoids, having scale lengths of 1.46/0.49/0.39 kpc, an angle
of the main axis with regards to the Sun-Galactic centre direction of 13\deg, and a total mass of  $6.1\times 10^{9}$\Msun. The age is  assumed to be 8 Gyr or more, and the mean metallicity is approximately solar. The second component is modelled by a triaxial exponential Ferrer ellipsoid less massive ($2.6\times 10^{8}$)\Msun, with scale lengths of 4.44/1.31/0.80 kpc, also old with a poorer metallicity of -0.35 dex. { A slight flare of the bar component is able to explain the double clumps qualitativeley  at latitudes higher than 6\deg}.

This model with two components explains the observed metallicity distribution function and its gradient along the minor axis by a transition between the two components dominating the counts, and also better reproduces the detailed distribution in the CMDs. A preliminary comparison with Brava data \cite{Rich2007} shows that the main component is a fast rotator, so  is probably a boxy bar (the object of a paper in preparation). The second structure (the thicker bulge) has a smaller velocity dispersion, thus could be either a classical bulge flattened by the potential of the bar or formed by early mergers, as in the \cite{Bournaud07} scenario. 
Testing of these scenarios are on-going, using our model and comparing predictions with kinematical and metallicity data in the whole bulge region.
 The forthcoming surveys, APOGEE project \cite{Majewski}, Gaia-ESO spectroscopic survey, among others, and later the ESA Gaia mission (http://www.rssd.esa.int/index.php?project=GAIA) are all expected to offer new insights into the kinematics and abundances of the central region of the Galaxy, giving strong constraints on any  scenario for the  formation of the bulge and bar stellar populations. We emphasize that our model can be used for preparing these future surveys and will be a useful tool for their interpretation.

\begin{acknowledgements}

This publication makes use of data products from the Two Micron All Sky Survey, which is a joint project of the 
University of Massachusetts and the Infrared Processing and Analysis Center/California Institute of Technology, 
funded by the National Aeronautics and Space Administration and the National Science Foundation. The CDSClient package was used for the remote querying of the 2MASS dataset. This material is based upon work supported in part by the National Science Foundation under Grant No. 1066293 and the hospitality of the Aspen Center for Physics. 
Simulations were executed on computers from the Utinam Institute of the Universit\'e de Franche-Comt\'e, supported by the R\'egion de Franche-Comt\'e and Institut des Sciences de l'Univers (INSU).
We acknowledge the French spatial agency, the Centre National d'Etude Spatiale, which has provided funding for D.J. Marshall, and the support of the french Agence Nationale de la Recherche under contract ANR-2010-BLAN-0508-01OTP. 
We thank Lia Athanassoula for fruitful discussions and Carine Babusiaux and Vanessa Hill for providing  their data and advices. 

\end{acknowledgements}
{}

\end{document}